\documentclass[a4paper,12pt]{report}

\usepackage{amssymb}
\usepackage{amsmath}

\newcommand{\nl}{\newline}
\newcommand{\beq}{\begin{equation}}
\newcommand{\eeq}{\end{equation}}
\newcommand{\nec}{\newcommand}
\nec{\cts}{conformal transformations }
\nec{\fourg}{g_{\alpha\beta}} 
\nec{\grad}{\bigtriangledown}
\nec{\fourr}{^{(4)}R}
\nec{\detg}{^{(4)}g}
\nec{\eins}{\Biggl(R_{\alpha\beta} - \frac{1}{2}g_{\alpha\beta}R\Biggr)}
\nec{\kk}{K^{ab}K_{ab}}
\nec{\bec}{\begin{center}}
\nec{\eec}{\end{center}}
\nec{\bb}{B^{ab}B_{ab}}
\nec{\rp}{R - 8\frac{\grad^{2}\psi}{\psi}}
\nec{\pipi}{\pi^{ab}\pi_{ab}}
\nec{\beqq}{\begin{equation*}}
\nec{\eeqq}{\end{equation*}}
\nec{\V}{V(\psi)}
\nec{\delg}{\frac{\partial g_{ab}}{\partial t}}
\nec{\wt}{\widetilde}
\nec{\gt}{\longrightarrow}
\nec{\wh}{\widehat}
\begin{document}

\null
\vspace{5mm}
\thispagestyle{empty}
\begin{center}
\vspace{10mm}
{\LARGE \bf{Gravity on Conformal Superspace}}
\vspace{20mm}

{\LARGE {Bryan Kelleher}}\\
\vspace*{35mm}

\end{center}
\vspace{12mm}

\begin{flushleft}
{\textsf{Thesis submitted in fulfillment of the requirements of the degree 
of}}\\ 
\large \textsc\bf {Doctor of Philosophy}\\
{\textsf{from the}}\\
{\textsf{Department of Physics}}\\
{\textsf{University College Cork}}\\
{\textsf{National University of Ireland, Cork}}\\
%{\textsf{June 2003}}\\
%\end{center}
\vspace{12mm}

{\large \textsf{Supervisor}:  
	Prof. Niall \'{O} Murchadha}\\

\end{flushleft}

\vspace{18mm}
\centerline{June 2003}
\newpage
\null
\vspace{60mm}
\thispagestyle{empty}
\begin{center}
\Large{To my family}
\end{center}
\newpage
\null
\vspace{0mm}

{\huge \bf{Acknowledgements}}\\
There are many people I wish to thank. Firstly, Niall, go raibh m\'{\i}le, 
m\'{\i}le maith agat. It has been both a pleasure and a privilege. Thanks to 
the entire physics department for everything over the years - great times, a 
superb atmosphere and lifelong memories. Last - but most \emph{definitely} not 
least - thanks to my parents, my brothers and sister, my wife to be Gill and my
extended family and friends. I could not (and more than likely would not) have
done it without you.
\thispagestyle{empty}
\newpage
\begin{abstract}
\noindent{The configuration space of general relativity is \emph{superspace} - 
the space of all Riemannian $3$-metrics modulo diffeomorphisms. However, it 
has been argued that the configuration space for gravity should be 
\emph{conformal superspace} - the space of all Riemannian $3$-metrics modulo
diffeomorphisms \emph{and} conformal transformations. Taking this conformal 
nature seriously leads to a new theory of gravity which although very similar 
to general relativity has some very different features particularly in 
cosmology and quantisation. It should reproduce the standard tests of general 
relativity. The cosmology is studied in some detail. The theory is incredibly 
restrictive and as a result admits an extremely limited number of possible 
solutions. The problems of the standard cosmology are addressed and most 
remarkably the cosmological constant problem is resolved in a natural way. The 
theory also has several attractive features with regard to quantisation 
particularly regarding the problem of time.}
\end{abstract}
\pagenumbering{roman}
\tableofcontents

\baselineskip=.65cm
\chapter{Introduction}
\pagenumbering{arabic}
\section{Introduction}
As formulated by Einstein, the natural arena for gravity as represented by 
general relativity (GR) is spacetime. We have a purely $4$-dimensional 
structure and the $4$-geometry reigns. (The invention of GR was a truly 
\emph{monumental} achievement and no offence is intended by any attempt here to
 suggest an alternative theory.) Dirac \cite{dir} and Arnowitt, Deser 
and Misner (ADM) \cite{adm} reformulated the theory in canonical form 
which is more in-keeping with other areas of modern physics. This formulation 
led to Wheeler's identification of the configuration space as superspace and GR
 as the theory of the evolution of the $3$-geometry which led to the coining 
(again by Wheeler) of geometrodynamics. To get superspace one first considers 
\emph{Riem} the space of all Riemannian $3$-geometries. Superspace is then 
\emph{Riem} modulo diffeomorphisms, that is, we identify all $3$-geometries 
related by diffeomorphisms.
\\ \\
York \cite{yor} went further and identified the conformal $3$-geometry with
 the dynamical degrees of freedom of the gravitational field. The correct 
configuration space for gravity should not be superspace but rather 
\emph{conformal superspace} - superspace modulo conformal transformations. 
Barbour and \'{O} Murchadha (BOM) \cite{bom} went further again and formulated
 a theory with conformal superspace at the very core.
\\ \\
We'll begin with a brief review of GR as found from the Einstein-Hilbert 
action and the ADM formulation. We'll then discuss the York approach and the 
original BOM theory. All of this will serve as a warm up 
(albeit, a necessary warm up) to the real focus of this work.
\section{General Relativity}
Although Einstein developed GR using beautiful physical reasoning and 
principles it is the Hilbert derivation from an action principle which is more 
instructive to us. (We will however refer to the action as the Einstein-Hilbert
 action as it was Einstein's work which inspired Hilbert to find the action to 
begin with.)
\\ \\
The Einstein-Hilbert action of general relativity is well known. It has the 
form
\beq \label{eh} S = \frac{1}{16\pi}\int \sqrt{-^{(4)}g}\: \fourr \: d^{4}x \eeq
 where $g_{\alpha\beta}$ is the 4-metric and $^{(4)}R$ is the four dimensional 
Ricci scalar. The action is varied with respect to $g_{\alpha\beta}$ and the 
resulting equations are the (vacuum) Einstein equations
\beq   G^{\alpha\beta} = \Biggl(R^{\alpha\beta} - 
\frac{1}{2}g^{\alpha\beta}R\Biggr) = 0   \eeq
matter sources may be included in the action and the resulting equations of 
motion are the full Einstein equations
\beq \label{eeqs} G^{\alpha\beta} = 8\pi T^{\alpha\beta} \eeq
where $T^{\alpha\beta}$ is the energy-momentum tensor of the matter field given
 by
\beq T^{\alpha\beta} = g^{\alpha\beta}L_{\text{matter}} - 
2\frac{\delta L_{\text{matter}}}{\delta g_{\alpha\beta}} \eeq
\section{(3+1)-Decomposition}
Before we consider the new theory it will be instructive to recall the ADM 
treatment of general relativity as much of this will carry straight over to the
new theory.
\\ \\
The idea in the ADM treatment is that a thin-sandwich $4$-geometry is 
constructed from two $3$-geometries separated by the proper time $d\tau$. The 
$4$-metric found from the ADM construction is
\begin{gather}
\begin{Vmatrix} ^{(4)}g_{00} & ^{(4)}g_{0k} \\ \\  ^{(4)}g_{i0} & ^{(4)}g_{ik} 
\end{Vmatrix} \quad
= \quad \begin{Vmatrix} (N^{s}N_{s} - N^{2}) & N_{k} \\ \\  N_{i} & g_{ik}
\end{Vmatrix}
\end{gather}
$N = N(t,x,y,z)$ is the lapse function given by
\beq d\tau = N(t,x,y,z)dt \eeq
and $N^{i}= N^{i}(t,x,y,z)$ are the shift functions given by
\beq x^{i}_{2}(x^{m}) = x^{i}_{1} - N^{i}(t,x,y,z)dt \eeq
where $x^{i}_{2}$ is the position on the ``later'' hypersurface corresponding 
to the position $x^{i}_{1}$ on the ``earlier'' hypersurface. The indices in the
 shift are raised and lowered by the $3$-metric $g_{ij}$.
\\ \\
The reciprocal $4$-metric is
\begin{gather}
\begin{Vmatrix} ^{(4)}g^{00} & ^{(4)}g^{0k} \\ \\  ^{(4)}g^{i0} & ^{(4)}g^{ik} 
\end{Vmatrix} \quad
= \quad \begin{Vmatrix}  -1/N^{2} & N^{k}/N^{2} \\ \\  N^{i}/N^{2} & 
g^{ik} - N^{i}N^{k}/N^{2}
\end{Vmatrix}
\end{gather}
The volume element has the form
\beq \sqrt{^{(4)}g}\;d^{4}x = N\sqrt{g}\;dt\;d^{3}x \eeq
This construction of the four metric also automatically determines the 
components of the unit timelike normal vector $\mathbf{\underline{n}}$. We get
\beq n_{\beta} = (-N,0,0,0) \eeq
and raising the indices using $^{(4)}g^{\alpha\beta}$ gives us
\beq n^{\alpha} = (1/N, -N^{m}/N) \eeq
\\ \\
Consider now the Einstein-Hilbert action
\beq S = \int \sqrt{-^{(4)}g}\: \fourr \: d^{4}x \eeq
Using the Gauss-Codazzi relations we get
\beq \fourr = R - (trK)^{2} + \kk - 2A^{\alpha}_{\;\; ;\alpha} \eeq 
where $A^{\alpha}$ is given by (as earlier)
\beq A^{\alpha} = \biggl(n^{\alpha}trK + a^{\alpha}\biggr) \eeq
$n^{\alpha}$ is the unit timelike normal and 
\beq a^{\alpha} = n^{\alpha}_{\;\; ;\beta}n^{\beta} \eeq
is the four-acceleration of an observer travelling along 
$\mathbf{\underline{n}}$. It is easily verified that $a^{0} = 0$ and that 
$a^{i} = \frac{\grad^{i}N}{N}$. Substituting into the action gives
\beq \label{admaction} S = \int N\sqrt{g}(R - (trK)^{2} + \kk) dtd^{3}x \eeq
where the total divergence $A^{\alpha}_{\;\; ;\alpha}$ has been discarded. 
$\mathbf{K}$ is the \emph{extrinsic curvature} given by
\beq \mathbf{K} = -\frac{1}{2}\mathbf{\pounds_{\underline{n}}}\mathbf{g} \eeq
the Lie derivative of the $3$-metric  $\mathbf{g}$ along 
$\mathbf{\underline{n}}$. In the coordinates we are using here the extrinsic 
curvature takes the form
\beq K_{ab} = -\frac{1}{2N}\biggl(\frac{\partial g_{ab}}{\partial t} - 
N_{a:b} - N_{b;a} \biggr) \eeq
The action is varied with respect to $\frac{\partial g_{ab}}{\partial t}$ to 
get the canonical momentum
\beq \pi^{ab} = \sqrt{g}\biggl(g^{ab}trK - K^{ab}\biggr) \eeq
and varied with respect to $N$ and $N_{a}$ to give the initial value equations
\beq \mathcal{H} = 0 \:\:\: \text{and} \:\:\: \mathcal{H}^{a} = 0 \eeq
respectively, where
\beq \label{ham} \mathcal{H} = \sqrt{g}\biggl(\pi^{ab}\pi_{ab} - 
\frac{1}{2}(tr\pi)^{2}\biggr) - \sqrt{g}R \eeq
and 
\beq \label{mom} \mathcal{H}^{a} = -2\pi^{ab}_{\;\;\;\; ;b} \eeq
these are known as the Hamiltonian constraint and the momentum constraint 
respectively. The Hamiltonian of the theory is then given by
\beq H = \int\biggl(N\mathcal{H} + N_{i}\mathcal{H}^{i}\biggr)\;\;d^{3}x \eeq
We get evolution equations for $g_{ab}$ and $\pi^{ab}$ by varying the 
Hamiltonian with respect to $\pi^{ab}$ and $g_{ab}$ respectively, using 
Hamilton's equations
\beq \delg = \frac{\delta H}{\delta\pi^{ab}} \eeq
\beq \frac{\partial\pi^{ab}}{\partial t} = - \frac{\delta H}{\delta g_{ab}} 
\eeq
These equations propagate the constraints. Solutions are a pair 
$\{g_{ab},\pi^{ab}\}$ which satisfy the constraints and are then evolved using 
the evolution equations. The lapse and shift ($N$ and $N^{i}$) are specified 
initially but after that are freely specifiable. This is the $4$-dimensional 
covariance.
\section{York's Approach}
The Hamiltonian and momentum constraints correspond to the $00$ and $0a$ 
components of Einstein's equations (\ref{eeqs}). They are equivalently 
initial-value constraints. We need to be able to find initial data which 
satisfy these. One method was proposed by Baerlein, Sharp and Wheeler (BSW) 
\cite{bsw}. This is known as the \emph{thin-sandwich conjecture}. First the 
pair $\{g_{ab},\delg\}$ are freely specified and then the momentum constraint 
is solved for the shift $N^{i}$. Although progress has been made, a regular 
method to solve this has not been found. A second method is York's conformal 
approach.
\\ \\
There are actually two different York methods (although they are very 
intimately related). In the first we begin with a maximal hypersurface. That 
is, the trace of the momentum is zero
\beq \label{max} tr\pi = 0 \eeq
everywhere on the hypersurface. Under these conditions the momentum constraint 
(\ref{mom}) is invariant under a conformal transformation of the form
\beq g_{ab} \gt \phi^{4}g_{ab} \eeq
\beq \pi^{ab} \gt \phi^{-4}\pi^{ab} \eeq
Transforming the Hamiltonian constraint under the same transformation gives the 
Lichnerowicz equation
\beq \pi^{ab}\pi_{ab}\phi^{-7} - R\phi + 8\grad^{2}\phi = 0 \eeq
York's approach is to solve the momentum constraint in a conformally invariant 
way (and such a way is well known) and then to solve the Lichnerowicz equation 
for $\phi$. The physical data is then $\{\phi^{4}g_{ab},\phi^{-4}\pi^{ab}\}$.
\\ \\
It turns out that the decoupling of the two constraints is still simple when 
the initial hypersurface has constant mean curvature (CMC) rather than being 
maximal. The CMC condition is that
\beq \label{cmc} trp = \frac{tr\pi}{\sqrt{g}} = \text{spatial constant} \eeq
We should introduce some new terminology here. The tracefree part of the 
momentum is
\beq \sigma^{ab} = \pi^{ab} - \frac{1}{3}g^{ab}tr\pi \eeq
Now, if the CMC condition holds then the momentum constraint reduces to
\beq \grad_{b}\sigma^{ab} = 0 \eeq
Now the tracefree part is transverse-traceless (TT). This property is invariant
 under the conformal transformation
\beq g_{ab} \gt \omega^{4}g_{ab} \eeq
\beq \sigma^{ab} \gt \omega^{-4}\sigma^{ab} \eeq
It is important here that $tr\pi$ now transforms in a different way to the 
tracefree part $\sigma^{ab}$. For the momentum constraint to be conformally invariant we need to define
\beq trp = \frac{tr\pi}{\sqrt{g}} \gt trp \eeq
That is, $trp$ transforms as a conformal scalar. Since there is a well known 
method to find a TT tensor we can find the pair $\{g_{ab},\sigma^{ab}\}$ 
easily. The Hamiltonian constraint transforms to become
\beq \sigma^{ab}\sigma_{ab}\phi^{-7} - \frac{1}{6}(tr\pi)^{2}\phi^{5} - R\phi +
 8\grad^{2}\phi = 0 \eeq
the extended Lichnerowicz equation. Specifying $g_{ab}$, $\sigma^{ab}$ and 
$trp$ we can solve for $\phi$ and then our physical data is 
$\{\phi^{4}g_{ab},\omega^{-4}\sigma^{ab},trp\}$.
\\ \\
In GR the conditions (\ref{max}) and (\ref{cmc}) are gauge conditions. If we 
are dealing with a manifold which is compact without boundary then we cannot 
have the maximal condition more than once. However, we may have the CMC 
condition holding always. It yields a foliation that is extremely convenient in
 the case of globally hyperbolic spacetimes. It is unique and the value of 
$trp$ increases monotonically either from $-\infty$ to $+\infty$ in n the case 
of a big bang to big crunch cosmological solution or from $-\infty$ to $0$ in 
the case of eternally expanding universes. In the first case the volume of the 
universe increases monotonically from $0$ to a point of maximum expansion at 
which the hypersurface is maximal. From this point on it decreases 
monotonically back to $0$. The volume cannot stay constant except (momentarily)
 at the maximum expansion when $trp = 0$. Thus, in GR the volume is dynamic 
which is of course the standard explanation of the cosmological redshift. One 
further point is that the quantity
\beq \label{yortim} \tau = \frac{2}{3}trp \eeq
is often interpreted as a notion of time, the York time, due to the properties 
of $trp$ noted above.
\subsection{Gauge Fixing in GR}
It is important to notice the difference between a single use of the CMC 
condition to find initial data and subsequent use of the condition when the 
data is propagated. This is by no means guaranteed. As noted earlier, once the 
initial data has been specified the lapse and shift are freely specifiable. To 
maintain the CMC slicing during the evolution it is necessary to choose the 
lapse in a particular way. Using the evolution equations we get
\beq \frac{\partial trp}{\partial t} = 2NR - 2\grad^{2}N + \frac{N(trp)^{2}}{4}
 + \frac{\grad_{c}tr\pi}{\sqrt{g}}N^{c} \eeq
To ensure CMC slicing we need to set $\grad_{c}tr\pi = 0$ and 
$\frac{\partial trp}{\partial t} = C$. That is
\beq \label{lf1} \frac{\partial trp}{\partial t} = 2NR - 2\grad^{2}N + 
\frac{N(trp)^{2}}{4} = C \eeq
where $C$ is a spatial constant but not necessarily a temporal constant. If we 
wish to maintain a maximal slicing we must have $tr\pi = 0$ and 
$\frac{\partial trp}{\partial t} = 0$. Thus
\beq \label{lf2} \frac{\partial trp}{\partial t} = 2NR - 2\grad^{2}N = 0 \eeq
As mentioned earlier, this particular condition cannot be maintained in a 
spacetime which is compact without boundary but can be maintained in an 
asymptotically flat spacetime. This will be dealt with in more detail later and
 will be of huge significance in the conformal theories developed. It should be
 noted that the conditions (\ref{lf1}) and (\ref{lf2}) do not fix the lapse 
uniquely since they are homogeneous in the lapse. They fix $N$ up to a global 
reparameterisation
\beq N \gt f(t)N \eeq
where $f(t)$ is an arbitrary monotonic function of $t$. These lapse-fixing 
equations arise naturally in the conformal theory as we shall see.
\\ \\
We should be ready now to move onto the original BOM conformal theory. We shall
 present it in a slightly different form however. The BOM action was of the 
Jacobi form. We shall derive a Lagrangian in the spirit of the traditional ADM 
$(3+1)$-dimensional action of GR.
\section{Lagrangian and Hamiltonian Formulations}
\subsection{The Lagrangian}
To begin with let's recall the ADM action of GR (\ref{admaction})
\beq S = \int N\sqrt{g}(R - (trK)^{2} + \kk) dtd^{3}x \eeq
To find the conformal action we simply transform the Lagrangian under the 
transformation
\beq g_{ab} \gt \psi^{4}g_{ab} \eeq
We need to define how the lapse and shift are transformed under such a 
transformation. In a later chapter we will see that this theory can be found 
using a $4$-dimensional action where
\beq g_{\alpha\beta} \gt \psi^{4}g_{\alpha\beta} \eeq
and under this we would have
\beq N \gt \psi^{2}N \eeq
and
\beq N_{i} \gt \psi^{4}N_{i} \eeq
Let's adopt these as our transformation rules. Under such a transformation
\beq R \gt \psi^{-4}\biggl(\rp\biggr) \eeq
The extrinsic curvature is a little more tricky. We have
\beq K_{ab} = -\frac{1}{2N}\biggl(\delg - (KN)_{ab}\biggr) \eeq
where
\beq (KN)_{ab} = \grad_{a}N_{b} + \grad_{b}N_{a} \eeq
Under a conformal transformation the various quantities behave as
\beq
\begin{split} N \gt & \psi^{2}N \\
\delg \gt & \psi^{4}\delg + 4\psi^{3}g_{ab}\frac{\partial\psi}{\partial t} \\
(KN)_{ab} \gt & \psi^{4}(KN)_{ab} + 4\psi^{3}g_{ab}N^{c}\grad_{c}\psi 
\end{split} \eeq
Thus
\beq K_{ab} \gt \psi^{2}B_{ab} = -\frac{\psi^{2}}{2N}\biggl(\delg - (KN)_{ab} -
 \theta g_{ab}\biggr) \eeq
where
\beq \theta = -\frac{4}{\psi}\biggl(\frac{\partial\psi}{\partial t} - 
N^{c}\grad_{c}\psi\biggr) \eeq
The Lagrangian is thus
\beq \label{admish} \mathcal{L} = N\sqrt{g}\psi^{4}\biggl(\rp + B^{ab}B_{ab} - 
(trB)^{2}\biggr) \eeq
\underline{Note}: We had $\theta$ in terms of $\psi$: $ \theta = 
-\frac{4}{\psi}\biggl(\dot{\psi} - \psi_{,i}N^{i}\biggr) $.
We can also find a coordinate independent form for $\mathbf{B}$. This is
\beq \mathbf{B} = 
-\frac{1}{2}\psi^{-4}\mathbf{\pounds_{\underline{n}}}(\psi^{4}\mathbf{g}) \eeq
This is analogous to the expression
\beq \mathbf{K} = -\frac{1}{2}\mathbf{\pounds_{\underline{n}}}(\mathbf{g}) 
\eeq
for the extrinsic curvature $\mathbf{K}$ in general relativity.
\subsection{Constraints and Evolution Equations}
We can perform the usual variations to find the constraints of the theory. 
Let's vary with respect to $N$ first. This gives us,
\beq \rp + (trB)^{2} - \bb = 0  \eeq
Varying with respect to $N^{a}$ gives us,
\beq \label{bab} \grad^{b}\biggl(\psi^{4}\biggl(g_{ab}trB - 
B_{ab}\biggr)\biggr) - 4\psi^{3}\psi_{,a}trB = 0  \eeq
As noted earlier we may vary with respect to 
$\psi$ and $\dot{\psi}$ independently. The $\dot{\psi}$ variation gives us,
\beq \label{trb} trB = 0  \eeq
This greatly simplifies equation (\ref{bab}) which now becomes
\beq \label{bab2} \grad^{b}\biggl(\psi^{4}B_{ab}\biggr) = 0 \eeq
The $\psi$ variation gives us,
\beq N\psi^{3}\biggl(R - 7\frac{\grad^{2}\psi}{\psi}\biggr) - 
\grad^{2}\biggl(N\psi^{3}\biggr) = 0  \eeq
where we have used the other constraints to simplify. The constraints may 
appear more familiar if we write them in terms of the canonical momentum rather
 than $B_{ab}$. We find the canonical momentum, $\pi^{ab}$ by varying the 
action with respect to $\frac{\partial g_{ab}}{\partial t}$. We get
\beq \pi^{ab} = \sqrt{g}\psi^{4}\biggl(g^{ab}trB - B^{ab}\biggr) \eeq
Then using equation (\ref{trb}) we get
\beq \pi^{ab} = -\sqrt{g}\psi^{4}B^{ab} \eeq
The constraints are then,
\beq \label{c1} \pipi - g\psi^{8}\biggl(\rp\biggr) = 0 \eeq
\beq \label{c2} \grad_{b}\pi^{ab} = 0  \eeq
\beq \label{c3} tr\pi = 0  \eeq
\beq \label{c4} N\psi^{3}\biggl(R - 7\frac{\grad^{2}\psi}{\psi}\biggr) - 
\grad^{2}\biggl(N\psi^{3}\biggr) = 0 \eeq
\\
Equation (\ref{c1}) corresponds to the Hamiltonian constraint of General 
Relativity. Equation (\ref{c2}) is the usual momentum constraint of general
relativity which represents diffeomorphism invariance. Equation (\ref{c3}) is 
new and represents conformal invariance. Our initial data consists of a pair 
$(g_{ab},\pi^{ab})$ which must satisfy equations (\ref{c2}) and (\ref{c3}). 
These are the initial value equations. Equation (\ref{c1}) is used to find the
``conformal field'' $\psi$ once we have specified the initial data. Equation 
(\ref{c4}) is a lapse-fixing equation which is used to determine $N$ 
throughout. We must check if these constraints are propagated under evolution.
\\ \\
The evolution equations are found in the usual way. They are
\beq \frac{\partial g_{ab}}{\partial t} = Ng^{-\frac{1}{2}}\psi^{-4}\pi_{ab} +
(KN)_{ab} - \theta g_{ab} \eeq
and
\beq
\begin{split} \frac{\partial\pi^{ab}}{\partial t} = & - 
\sqrt{g}N\psi^{4}\biggl(R^{ab} - g^{ab}\biggl(\rp\biggr)\biggr) - 
2Ng^{-\frac{1}{2}}\psi^{-4}\pi^{ac}\pi^{b}_{\;\; c} \\ & + 
\sqrt{g}\psi\biggl(\grad^{a}\grad^{b}(N\psi^{3}) - 
g^{ab}\grad^{2}(N\psi^{3})\biggr) \\ & 
N\sqrt{g}\psi^{3}\biggl(\grad^{a}\grad^{b}\psi + 3g^{ab}\grad^{2}\psi\biggr)
\\ & + 4\sqrt{g}g^{ab}\grad_{c}(N\psi^{3})\grad^{c}\psi - 
6\sqrt{g}\grad^{(a}(N\psi^{3})\grad^{b)}\psi \\ & + 
\grad_{c}\biggl(N^{c}\pi^{ab}\biggr) - 
\pi^{bc}\grad_{c}N^{a} - \pi^{ac}\grad_{c}N^{b} - \theta\pi^{ab} 
\end{split} \eeq
It can be verified that these equations do indeed preserve the constraints.
\\ \\
We can see how similar the results are to those in York's approach. The 
Hamiltonian constraint has become the Lichnerowicz equation. The momentum is 
TT. Also, the lapse fixing equation is the gauge requirement of GR to preserve 
the $tr\pi = 0$ constraint. Of course, those equations are all secondary in GR 
whereas here they have arisen directly through a variational procedure!
\subsection{The Hamiltonian}
Now that we have found the momentum it is straightforward to find the 
Hamiltonian. As usual we have
\beq H = \int\biggl(\pi^{ab}\delg - \mathcal{L}\biggr)\;\;d^{3}x \eeq
We must write $\mathcal{L}$ in terms of the momentum $\pi^{ab}$. This is
\beq \mathcal{L} = N\sqrt{g}\psi^{4}\biggl(\rp + \frac{\pi^{ab}\pi_{ab} - 
\frac{1}{2}(tr\pi)^{2}}{g\psi^{8}}\biggr)\;\;d^{3}x \eeq
This leads to
\beq H = \int \Biggl[N\biggl[\frac{1}{\sqrt{g}\psi^{4}}\biggl(\pi^{ab}\pi_{ab} - 
\frac{1}{2}(tr\pi)^{2}\biggr) - \sqrt{g}\psi^{4}\biggl(\rp\biggr)\biggr] - 
2N_{a}\grad_{b}\pi^{ab} + \theta tr\pi\Biggr]\;\;d^{3}x \eeq
Recalling the constraints we see that yet again, as found by Dirac and ADM, the
Hamiltonian is a sum of the constraints with Lagrange multipliers.
\section{Jacobi Action}
Baerlein, Sharp and Wheeler \cite{bsw} constructed a Jacobi Action for general 
relativity.
Their action was,
\beq \label{bsw} S = \underline{+}\int 
d\lambda\int\sqrt{g}\sqrt{R}\sqrt{T_{\text{GR}}}d^{3}x \eeq
where
\beq T_{GR} = \biggl(g^{ac}g^{bd} - g^{ab}g^{cd}\biggr)\biggl(\frac{\partial 
g_{ab}}{\partial t} - (KN)_{ab}\biggr)\biggl(\frac{\partial 
g_{cd}}{\partial t} - (KN)_{cd}\biggr) \eeq
Variation with respect to $\frac{\partial g_{ab}}{\partial t}$ gives 
\beq \pi^{ab} = \sqrt{\frac{gR}{T_{GR}}}\biggl(g^{ac}g^{bd} - 
g^{ab}g^{cd}\biggr)\biggl(\frac{\partial g_{cd}}{\partial t} - (KN)_{cd}\biggr)
\eeq
This expression is squared to give the Hamiltonian constraint. The variation 
with respect to $N_{a}$ gives the momentum constraint. The evolution equations
are found in the usual way. The equations found with the Jacobi action are 
those of general relativity if we identify $2N$ and $\sqrt{\frac{T}{R}}$. We 
want to construct the analogous case in conformal gravity. Let us return to our
(3+1) Lagrangian,
\beq \mathcal{L} = N\sqrt{g}\psi^{4}\biggl(\rp - (trB)^{2} + \bb\biggr) \eeq
We can write this as 
\beq \mathcal{L} = \sqrt{g}\psi^{4}\biggl[N\biggl(\rp\biggr) + 
\frac{1}{4N}\biggl(\beta^{ab}\beta_{ab} - (tr\beta)^{2}\biggr)\biggr] \eeq
where $ \beta_{ab} = - 2NB_{ab} = \biggl(\frac{\partial g_{ab}}{\partial t} - 
(KN)_{ab} - \theta g_{ab}\biggr)$.
We now extremise with respect to $N$. This gives us,
\beq \label{Ntr} N = \underline{+}\frac{1}{2}\biggl(\beta^{ab}\beta_{ab} - 
(tr\beta)^{2}\biggr)^{\frac{1}{2}}\biggl(\rp\biggr)^{-\frac{1}{2}} \eeq
Substituting this back into the action gives us
\beq S = \underline{+}\int d\lambda\int\sqrt{g}\psi^{4}\sqrt{\rp}\sqrt{T}d^{3}x
\eeq
where $ T = \biggl(\beta^{ab}\beta_{ab} - (tr\beta)^{2}\biggr) $.
This is the conformal gravity version of the BSW action (\ref{bsw}).
\\ \\
We can do all the usual variations here: $N^{a}$, $\dot{\psi}$ and $\psi$. 
These give the momentum constraint, the conformal constraint and the 
lapse-fixing equation respectively. Because of the independent variations of 
$\dot{\psi}$ and $\psi$, it turns out that we may vary with respect to $\theta$
 and $\psi$ to get the conformal constraint and the lapse-fixing equation 
respectively. When we find the canonical momentum $\pi^{ab}$ we can ``square'' 
it to give the ``Hamiltonian constraint.''
\\ \\
Actually, this is precisely the BOM action found by starting with the BSW 
action and conformalising it under conformal transformations of the $3$-metric
\beq g_{ab} \longrightarrow \psi^{4}g_{ab} \eeq
The Jacobi action is manifestly 3-dimensional and its configuration space is 
naturally conformal superspace - the space of all 3-D Riemannian metrics modulo
diffeomorphisms and conformal rescalings.
\section{Conformally Related Solutions}
In conformal superspace conformally related metrics are equivalent. Thus 
conformally related solutions of the theory must be physically equivalent and 
so it is crucial that we have a natural way to relate such solutions. Suppose 
we have one set of initial data $ (g_{ab},\pi^{ab}) $. These must satisfy the 
constraints (\ref{c2}) and (\ref{c3}). We solve the Hamiltonian constraint 
(\ref{c1}) for our ``conformal field'' $\psi$. Suppose now we start with a 
different pair $(h_{ab},p^{ab})$ where $h_{ab} = \alpha^{4} g_{ab}$ and 
$\rho^{ab} = \alpha^{-4}\pi^{ab}$. Our new initial data is conformally related
 to the original set of initial data. This is allowed as 
``transverse-traceless''-ness is conformally invariant and so our initial data 
constraints are satisfied. All we must do is solve the new Hamiltonian 
constraint for our new conformal field $\chi$ say. This constraint is now
\beq \rho^{ab}\rho_{ab} = h\chi^{8}\biggl(R_{h} - 
8\frac{\grad_{h}^{2}\chi}{\chi}\biggr) \eeq
The subscript $h$ on $R$ and $\grad$ is because we are now dealing with the new
metric $h_{ab}$. We now solve this for $\chi$. It can be shown that we must 
have $\chi = \frac{\psi}{\alpha}$. That is, $\psi$ is automatically transformed
when our initial data is transformed.
\\ \\
Now,
\beq  \chi^{4}h_{ab} = \frac{\psi^{4}}{\alpha^{4}}\alpha^{4}g_{ab} = 
\psi^{4}g_{ab} \eeq
and
\beq \chi^{-4}\rho^{ab} = \psi^{-4}\pi^{ab} \eeq
If we label these as $ \widetilde{\,g_{ab}} = \psi^{4}g_{ab} $ and 
$ \widetilde{\,\pi^{ab}} = \psi^{-4}\pi^{ab} $ than we can write our 
constraints as
\beq \widetilde{\,\pi^{ab}}\widetilde{\,\pi_{ab}} = \widetilde{g}\widetilde{R}
\eeq
\beq \widetilde{\,\grad_{_{b}}}\widetilde{\, \pi^{ab}} = 0 \eeq
\beq \widetilde{tr\pi} = 0 \eeq
\beq \widetilde{N}\widetilde{R} - \widetilde{\,\grad^{2}}\widetilde{N} = 0 
\eeq
All conformally related solutions are identical in this form and thus we shall 
refer to this as the physical representation. The momentum constraint is 
identical in the two theories. The Hamiltonian constraint of general relativity
 on a maximal slice is identical to that here. The lapse-fixing equation in 
this representation looks just like the maximal slicing equation of general 
relativity. In this representation the evolution equations are
\beq \frac{\partial g_{ab}}{\partial t} = Ng^{-\frac{1}{2}}\pi_{ab} + (KN)_{ab}
\eeq
and
\beqq \frac{\partial\pi^{ab}}{\partial t} = - \sqrt{g}N\biggl(R^{ab} 
- g^{ab}R\biggr) + 
\grad^{a}\grad^{b}\biggl(\sqrt{g}N\biggr) \eeqq
\beqq - \sqrt{g}g^{ab}\grad^{2}N - 
2Ng^{-\frac{1}{2}}\pi^{ac}\pi^{b}_{\;\; c} \eeqq
\beq - \grad_{c}N^{a}\pi^{bc} - \grad_{c}N^{b}\pi^{ac} + 
\grad_{c}\biggl(N^{c}\pi^{ab}\biggr) \eeq
These are exactly those of general relativity on a maximal slice. Thus, 
solutions of general relativity in maximal slicing gauge are also solutions 
here. There are of course solutions of general relativity which do not have a 
maximal slicing and these are not solutions of the conformal theory.
\section{Topological Considerations}
So far we have not considered any implications which the topology of the 
manifold may have. In an asymptotically flat case we have no problems with the
theory as it stands. This is not the case however in a topology which is
compact without boundary.
\subsection{Integral Inconsistencies}
Recall the lapse-fixing equation of the theory in the physical representation 
(removing the ``hats'' for simplicity),
\beq NR - \grad^{2}N = 0 \eeq
Let's integrate this equation:
\beq \int\sqrt{g}NR\;\;d^{3}x - \int \sqrt{g}\grad^{2}N\;\;d^{3}x = 0 \eeq
The second term integrates to zero and so we just have
\beq \int \sqrt{g}NR\;\;d^{3}x = 0 \eeq
and so we must have that $N$ is sometimes positive and sometimes negative or 
else is identically zero. Suppose the first of these possibilities is true. 
Let's now restrict our integration of the lapse-fixing equation to the positive
 values of $N$ only. This has a real boundary, namely, $N = 0$. We thus have
\beq \int\sqrt{g}NR\;\;d^{3}x - \int \sqrt{g}\grad^{2}N\;\;d^{3}x = 0 \eeq
again. Now, the first integral
\beq \int\sqrt{g}NR\;\;d^{3}x \eeq
is positive definite. The second integral is
\beq - \int \sqrt{g} \grad^{2}N\;\; d^{3}x \eeq
which becomes a surface integral after integrating by parts
\beq - \int \sqrt{g}N\grad^{c}N\;\;d\Sigma_{c} \eeq
where $\Sigma_{c}$ is the boundary on which $N = 0$. Since $N$ is decreasing on
 the boundary we have that this term is positive definite. This means however 
that we have a vanishing sum of two positive definite quantities. This is a 
contradiction. Thus we must have $N \equiv 0$. We get frozen dynamics. (This is
 not the case with a manifold which is asymptotically flat so the earlier 
analysis works in that case.) Frozen dynamics also arises in general relativity
if one imposes a fixed$tr\pi = 0$ gauge condition. However, this is a problem 
of the gauge rather than a problem of the theory as with conformal gravity. 
(See \cite{abfom} for a treatment of this problem.)
\\ \\
The easiest way to resolve this problem involves a slight change to the action.
We introduce a volume term. The inspiration for this term comes from the Yamabe
theorem. The action is
\beq S = \int\frac{N\sqrt{g}\psi^{4}}{V^{2/3}}\biggl(\rp +B^{ab}B_{ab} - 
(trB)^{2}\biggr)\;\; d^{3}x\;\;dt \eeq
The $V$ in the denominator is the volume
\beq V = \int\sqrt{g}\psi^{6}\: d^{3}x \eeq
The power of $\frac{2}{3}$ on the volume leaves the action homogeneous in both 
$\psi$ and $g$. The constraints arising from this action are not very different
from the original constraints. We get firstly that
\beq \pi^{ab} = \frac{\sqrt{g}\psi^{4}}{V^{2/3}}\biggl(g^{ab}trB - 
B^{ab}\biggr) \eeq
The constraints are then
\beq \pipi = \frac{g\psi^{8}}{V^{4/3}}\biggl(\rp\biggr) \eeq
\beq \grad_{b}\pi^{ab} = 0  \eeq
\beq tr\pi = 0  \eeq
\beq N\psi^{3}\biggl(R - 7\frac{\grad^{2}\psi}{\psi}\biggr) - 
\grad^{2}\biggl(N\psi^{3}\biggr) = C\psi^{5} \eeq
The term $C$ is given by 
\beq \label{cvol} C = \int\frac{N\sqrt{g}\psi^{4}}{V}\biggl(\rp\biggr)\;\;
d^{3}x \eeq 
which arises due to the variation of the volume. In the physical representation
 these constraints become
\beq \pipi = \frac{gR}{\V^{4/3}} \eeq
\beq \grad_{b}\pi^{ab} = 0 \eeq
\beq tr\pi = 0 \eeq
\beq NR - \grad^{2}N = C \eeq
where $C$ is now
\beq  C = \frac{1}{2}\int \frac{\sqrt{g}\sqrt{R}\sqrt{T}}{V} d^{3}x = 
\Big<NR\Big> \eeq
\\ \\
\underline{Note}: $\Big<A\Big>$ is the average of $A$ given by the usual notion
of average 
\beq \Big<A\Big> = \frac{\int \sqrt{g}Ad^{3}x}{\int \sqrt{g} d^{3}x} \eeq
\\ \\
In this form, the lapse-fixing equation looks just like the constant mean 
curvature slicing equation of general relativity on a maximal slice. To check 
if we have any inconsistency this time, we integrate our lapse-fixing equation 
again. We need,
\beq \int\sqrt{g}NR\:d^{3}x - \int\grad^{2}N\:d^{3}x - 
\int \sqrt{g}C\:d^{3}x = 0 \eeq
This becomes 
\beq \int \sqrt{g}NR\:d^{3}x - \int \sqrt{g} 
\Big<NR\Big>\:d^{3}x = 0 \eeq
removing the second term which integrates to zero. The left hand side is then 
just
\beqq V\Big<NR\Big> - V\Big<NR\Big> \eeqq
\beq = 0 \eeq
as required. Thus, with the introduction of the volume term, we have removed 
the problem.
\\ \\
\underline{Note}: Although we have only used the physical representation 
in our integral tests it can be verified easily that everything also works out
in the general representation. Of course, in EVERY situation, this \emph{must} 
be true. We are losing \emph{nothing} by working in the physical 
representation.
\\ \\
We should consider the evolution equations again now that we have changed the 
action. The evolution equations become
\beq \frac{\partial g_{ab}}{\partial t} = 
Ng^{-\frac{1}{2}}V(\psi)^{\frac{2}{3}}\psi^{-4}\pi_{ab} + (KN)_{ab} + 
\theta g_{ab} \eeq
and
\beq
\begin{split} \frac{\partial\pi^{ab}}{\partial t} = & - 
\frac{N\sqrt{g}\psi^{4}}{V^{2/3}}\biggl(R^{ab} - g^{ab}\biggl(\rp\biggr)\biggr)
 - \frac{2NV^{4/3}}{\sqrt{g}\psi^{4}}\pi^{ac}\pi^{b}_{\;\; c} \\ & + 
\frac{\sqrt{g}\psi}{V^{2/3}}\biggl(\grad^{a}\grad^{b}(N\psi^{3}) - 
g^{ab}\grad^{2}(N\psi^{3})\biggr) \\ & + 
\frac{N\sqrt{g}\psi^{3}}{V^{2/3}}\biggl(\grad^{a}\grad^{b}\psi + 
3g^{ab}\grad^{2}\psi\biggr)
\\ & + \frac{4\sqrt{g}g^{ab}}{V^{2/3}}\grad_{c}(N\psi^{3})\grad^{c}\psi - 
\frac{6\sqrt{g}}{V^{2/3}}\grad^{(a}(N\psi^{3})\grad^{b)}\psi \\ & + 
\grad_{c}\biggl(N^{c}\pi^{ab}\biggr) - 
\pi^{bc}\grad_{c}N^{a} - \pi^{ac}\grad_{c}N^{b} - \theta\pi^{ab} - 
\frac{2}{3}\frac{\sqrt{g}g^{ab}C\psi^{5}}{V^{2/3}} \end{split} \eeq
where $C$ is as in (\ref{cvol}). As usual we can write these in the physical 
representation. In this form the evolution equations are
\beq \frac{\partial g_{ab}}{\partial t} = 
Ng^{-\frac{1}{2}}V^{\frac{2}{3}}\pi_{ab} + (KN)_{ab} \eeq
and
\beq
\begin{split} \frac{\partial\pi^{ab}}{\partial t} = & - 
\frac{N\sqrt{g}}{V^{2/3}}\biggl(R^{ab} - g^{ab}R\biggr)
 - \frac{2NV^{4/3}}{\sqrt{g}}\pi^{ac}\pi^{b}_{\;\; c} \\ & + 
\frac{\sqrt{g}}{V^{2/3}}\biggl(\grad^{a}\grad^{b}N - 
g^{ab}\grad^{2}N\biggr) + \grad_{c}\biggl(N^{c}\pi^{ab}\biggr) \\ & - 
\pi^{bc}\grad_{c}N^{a} - \pi^{ac}\grad_{c}N^{b} - 
\frac{2}{3}\frac{\sqrt{g}g^{ab}C}{V^{2/3}} \end{split} \eeq
where now $C = \Big<NR\Big>$. 
\\ \\
Let's define
\beq \wh{\pi^{ab}} = V^{2/3}\pi^{ab} \eeq
Then the constraints are
\beq \wh{\pi^{ab}}\wh{\pi_{ab}} - gR = 0 \eeq
\beq \grad_{b}\wh{\pi^{ab}} = 0 \eeq
\beq \wh{tr\pi} = 0 \eeq
The lapse-fixing equation is
\beq NR - \grad^{2}N = C \eeq
These are precisely the constraints and gauge fixing conditions for propagated 
maximal slicing in GR. The evolution equations are
\beq \frac{\partial g_{ab}}{\partial t} = 
Ng^{-\frac{1}{2}}\wh{\pi_{ab}} + (KN)_{ab} \eeq
and
\beq
\begin{split} \frac{\partial\wh{\pi^{ab}}}{\partial t} = & - 
N\sqrt{g}\biggl(R^{ab} - g^{ab}R\biggr)
 - \frac{2N}{\sqrt{g}}\wh{\pi^{ac}}\wh{\pi^{b}_{\;\; c}} \\ & + 
\sqrt{g}\biggl(\grad^{a}\grad^{b}N - g^{ab}\grad^{2}N\biggr) + 
\grad_{c}\biggl(N^{c}\wh{\pi^{ab}}\biggr) \\ & - 
\wh{\pi^{bc}}\grad_{c}N^{a} - \wh{\pi^{ac}}\grad_{c}N^{b} - 
\frac{2}{3}\frac{\sqrt{g}g^{ab}C}{V^{2/3}} \end{split} \eeq
which are identical to those in GR apart from the global $C$ term in the 
equation for $\pi^{ab}$.
\\ \\
We can easily find the Hamiltonian and the Jacobi action for the new form. 
They are
\beq H = \int\biggl[N\frac{V^{2/3}}{\sqrt{g}\psi{4}}\biggl(\pi^{ab}\pi_{ab} 
- \frac{1}{2}(tr\pi)^{2} - 
\frac{\sqrt{g}\psi^{4}}{V^{2/3}}\biggl(\rp\biggr)\biggr)\biggr]\;\;d^{3}x 
\eeq
and
\beq S = \int\: d\lambda\int\frac{\sqrt{g}\psi^{4}\sqrt{\rp}\sqrt{T}}
{V(\psi)^{\frac{2}{3}}}\: d^{3}x \eeq
Note again the homogeneity throughout in $\psi$.
\section{Other Results}
There has been work on other aspects of this theory not described here. It 
is unnecessary from the point of view of this work while, of course, being 
valuable in itself with a number of worthwhile results most notably on the 
constraint algebra and the Hamilton-Jacobi theory. The interested reader can
 find this in \cite{bk}.
\section{Problem}
Although the theory has emerged beautifully and easily form very natural 
principles we can find at least one major problem immediately. Consider the 
volume of a hypersurface $V$
\beq V = \int\sqrt{g}\;\;d^{3}x \eeq
Taking the time derivative of this we get
\beq \frac{\partial V}{\partial t} = \int\frac{1}{2}\sqrt{g}g^{ab}\delg\;\;
d^{3}x \eeq
This becomes
\beq \frac{\partial V}{\partial t} = \int \biggl(NV^{2/3}tr\pi + 
2\grad_{c}N^{c}\biggr)\;\;d^{3}X \eeq
Now since $tr\pi = 0$ the entire expression is zero. That is
\beq \frac{\partial V}{\partial t} = 0 \eeq
and the volume of the universe is static. This rules out expansion and thus the
standard cosmological solution is lost. In particular, the red-shift, an 
experimental fact, is unexplained. This is a serious shortcoming. All is not 
lost however...
\chapter{A New Hope}
\section{The Need For A Change}
Despite all the promising features of the theory there is at least one major 
drawback. We can find the time derivative of the volume quite easily and get 
that it is proportional to $tr\pi$ and thus is zero. That is, the volume does 
not change and so the theory predicts a static universe and we cannot have 
expansion. This is quite a serious problem as the prediction of expansion in GR
 is considered to be one of the theory's greatest achievements. We are left 
with the following options:\nl
\\
(a) Abandon the theory; \nl
(b) Find a new explanation of the red-shift (among other things); \nl
(c) Amend the theory to recover expansion. \nl
\\
The first option seems quite drastic and the second, while certainly the most 
dramatic, also seems to be the most difficult. Thus, let's check what we can 
find behind door (c).
\subsection{Resolving The problem(s)}
Any change to the theory needs to be made at the level of the Lagrangian and so
 we'll return to our earlier expression for $\mathcal{L}$
\beq \mathcal{L} = N\sqrt{g}\psi^{4}\biggl(\rp + B_{ab}B^{ab} - 
(trB)^{2}\biggr) \eeq
but naively change the form of $B_{ab}$ to
\beq \label{newb} B_{ab} = - \frac{1}{2N}\biggl(\delg - (KN)_{ab} - 
\grad_{c}\xi^{c}g_{ab}\biggr) \eeq
Let's vary the action with respect to $\xi^{c}$. We get
\beq \begin{split} \delta\mathcal{L} = &
N\sqrt{g}\psi^{4}\biggl(2B^{ab} - 2trBg^{ab}\biggr)\delta B_{ab} \\ = & 
2N\sqrt{g}\psi^{4}\biggl(B^{ab} - 2trBg^{ab}\biggr)
\biggl(\frac{-1}{2N}\biggr)\biggl(-\grad_{c}\delta\xi^{c}g_{ab}\biggr) \\ = & 
-2\sqrt{g}\psi^{4}trB\grad_{c}\delta\xi^{c} \end{split} \eeq
Integrating by parts gives
\beq \delta\mathcal{L} = 2\sqrt{g}\grad_{c}(trB\psi^{4})\delta{\xi^{c}} \eeq
and so
\beq \grad_{c}(trB\psi^{4}) = 0 \eeq
Recall that we had
\beq tr\pi = \psi^{4}trB \eeq
and so the constraint is
\beq \grad_{c}tr\pi = 0 \eeq
the CMC condition.
\\
However, since we have the same form for $\mathcal{L}$ as before our 
lapse-fixing equation is unchanged and as a result, the constraint is not 
propagated unless $tr\pi = 0$. Thus we haven't gained anything. We need a 
further change.
\\
It will prove instructive to split $B_{ab}$ into its trace and tracefree parts.
(The reason for this will become clear quite soon.) We label the tracefree part
 as $S_{ab}$. Thus we have
\beq B_{ab} = S_{ab} + \frac{1}{3}g_{ab}trB \eeq
We shall retain the new form of $B_{ab}$ as defined above in (\ref{newb}) all 
the same. The Lagrangian now reads
\beq \mathcal{L} = N\sqrt{g}\psi^{4}\biggl(\rp + S_{ab}S^{ab} - 
\frac{2}{3}(trB)^{2}\biggr) \eeq
We still need to make one further change. We'll simply stick in an additional 
$\psi$ term to the $trB$ part. (This is equivalent to redefining our conformal 
transformation so that $S_{ab}$ and $trB$ transform in different ways.) The 
Lagrangian takes the form
\beq \mathcal{L} = N\sqrt{g}\psi^{4}\biggl(\rp + S_{ab}S^{ab} - 
\frac{2}{3}\psi^{n}(trB)^{2}\biggr) \eeq
\\
Before we continue, one interesting point about $S_{ab}$ is the following. We 
have
\beq S_{ab} = B_{ab} - \frac{1}{3}g_{ab}trB \eeq
Let's write this out explicitly. We have
\beq S_{ab} = - \frac{1}{2N}\biggl(\delg - (KN)_{ab} - 
\grad_{c}\xi^{c}g_{ab}\biggr) - \frac{1}{3}g_{ab}\biggl(g^{cd}\frac{\partial 
g_{cd}}{\partial t} - g^{cd}(KN)_{cd} - 3\grad_{c}\xi^{c}\biggr) \eeq
Splitting this up further gives
\beq S_{ab} = - \frac{1}{2N}\biggl(\delg - (KN)_{ab}\biggr) + 
\frac{3}{2N}g_{ab}\grad_{c}\xi^{c} - 
\frac{1}{3}g_{ab}\biggl(g^{cd}\frac{\partial g_{cd}}{\partial t} - 
g^{cd}(KN)_{cd}\biggr) -\frac{3}{2N}g_{ab}\grad_{c}\xi^{c} \eeq
and with a simple cancellation
\beq S_{ab} = - \frac{1}{2N}\biggl(\delg - (KN)_{ab}\biggr) - 
\frac{1}{3}g_{ab}\biggl(g^{cd}\frac{\partial g_{cd}}{\partial t} - 
g^{cd}(KN)_{cd}\biggr) \eeq
Of course, this is
\beq S_{ab} = K_{ab} - \frac{1}{3}g_{ab}trK \eeq
That is, $S_{ab}$ is the tracefree part of the extrinsic curvature and is 
\emph{independent} of \emph{any} conformal fields.
\\ \\
Let us find $\pi^{ab}$. This is done as usual by varying with respect to 
$\delg$. We get
\beq \begin{split} \delta\mathcal{L} & =  
2N\sqrt{g}\psi^{4}\biggl(2S^{ab}\delta S_{ab} - 
\frac{4}{3}\psi^{n}trBg^{ab}\delta B_{ab}\biggr) 
\\& = 2N\sqrt{g}\psi^{4}\biggl(S^{ab}\biggl(\delta B_{ab} - 
\frac{1}{3}g_{ab}g^{cd}\delta B_{cd}\biggr) - 
\frac{2}{3}\psi^{n}trBg^{ab}\delta B_{ab}\biggr)
\\& = 2N\sqrt{g}\psi^{4}\biggl(S^{ab} - 
\frac{2}{3}\psi^{n}trBg^{ab}\biggr)\delta B_{ab}
\\& = - \sqrt{g}\psi^{4}\biggl(S^{ab} - \frac{2}{3}\psi^{n}S^{ab}trB\biggr)
\delta \delg \end{split} \eeq
Thus,
\beq \pi^{ab} = -\sqrt{g}\psi^{4}S^{ab} + 
\frac{2}{3}\sqrt{g}\psi^{n+4}g^{ab}trB \eeq
Splitting $\pi^{ab}$ into its trace and tracefree parts will further clear 
things up. We'll label the split as
\beq \pi^{ab} = \sigma^{ab} + \frac{1}{3}g^{ab}tr\pi \eeq
Thus the tracefree part of $\pi^{ab}$ is
\beq \sigma^{ab} = -\sqrt{g}\psi^{4}S^{ab} \eeq
and the trace is given by
\beq tr\pi = 2\psi^{n+4}trB \eeq
Note that our value of $n$ is undefined as yet.
\\ \\
The constraints are found by varying with respect to $\xi^{c}$, $\psi$, $N$ and
$N^{a}$. The conformal constraint and the lapse-fixing equation are given by 
varying with respect to $\xi^{c}$ and $\psi$ respectively. These give
\beq \grad_{c}tr\pi = 0 \eeq
and
\beq N\psi^{3}\biggl(R - 7\frac{\grad^{2}\psi}{\psi}\biggr) - 
\grad^{2}(N\psi^{3}) + \frac{(trp)^{2}\psi^{7}}{4} = 0 \eeq
respectively. From the variation with respect to $N$ we get
\beq S_{ab}S^{ab} - \frac{2}{3}\psi^{n}(trB)^{2} - g\psi^{8}\biggl(\rp\biggr) =
 0 \eeq
which in terms of the momentum is
\beq \label{hami} \sigma_{ab}\sigma^{ab} - \frac{1}{6}\psi^{-n}(tr\pi)^{2} - 
g\psi^{8}\biggl(\rp\biggr) = 0 \eeq
and finally, from the variation with respect to $N^{a}$ we get
\beq \label{xxx} \grad_{b}\pi^{ab} = 0 \eeq
We require conformal invariance in our constraints. Under what conditions is 
the momentum constraint (\ref{xxx}) invariant? The tracefree part of the 
momentum, $\sigma^{ab}$, has a natural weight of $-4$ (from the original 
theory). That is
\beq \sigma^{ab} \gt \omega^{-4}\sigma^{ab} \eeq
If $tr\pi = 0$ then we have conformal invariance. If not however, we require 
various further conditions. We need
\beq \grad_{b}\sigma^{ab} = 0 \eeq
\beq \grad_{c}tr\pi = 0 \eeq
and that
\beq trp = \frac{tr\pi}{\sqrt{g}} \gt trp \eeq
under a conformal transformation. In our theory we have the first two 
conditions emerging directly and naturally from the variation. Thus we simply 
define $trp$ to transform as a conformal scalar as required. With this done our
 momentum constraint is conformally invariant.
\\ \\
Transforming the constraint (\ref{hami}) gives
\beq \sigma_{ab}\sigma^{ab} - \frac{1}{6}\psi^{-n}g(trp)^{2}\omega^{12 + n} - 
g\psi^{8}\biggl(\rp\biggr) = 0 \eeq
and so we must have $n = -12$ for conformal invariance.
The constraint then becomes
\beq \sigma_{ab}\sigma^{ab} - \frac{1}{6}\psi^{12}g(trp)^{2} - 
g\psi^{8}\biggl(\rp\biggr) = 0 \eeq
\\ \\
(Note: This is \emph{exactly} the Lichnerowicz equation from GR. However, we 
have found it directly from a variational procedure.)
\\ \\
Thus we have determined the unique value of $n$ and our constraints are
\beq \label{c1new} \sigma_{ab}\sigma^{ab} - \frac{1}{6}\psi^{12}(tr\pi)^{2} - 
g\psi^{8}\biggl(\rp\biggr) = 0 \eeq
\beq \label{c2new} \grad_{b}\pi^{ab} = 0 \eeq
\beq \label{c3new}\grad_{c}tr\pi = 0 \eeq
\beq \label{c4new} N\psi^{3}\biggl(R - 7\frac{\grad^{2}\psi}{\psi}\biggr) - 
\grad^{2}(N\psi^{3}) + \frac{(trp)^{2}\psi^{7}}{4} = 0 \eeq
\\ \\
Let's proceed to the Hamiltonian formulation.
\section{The Hamiltonian Formulation}
The earlier expression for $\pi^{ab}$ can be inverted to get $\delg$. We get
\beq \delg = \frac{2N}{\sqrt{g}\psi^{4}}\biggl(\sigma_{ab} - 
\frac{1}{6}g_{ab}tr\pi\psi^{12}\biggr) + (KN)_{ab} + 
g_{ab}\grad_{c}\xi^{c} \eeq
The Hamiltonian may then be found in the usual way. We get
\beq \mathcal{H} = 
\frac{N}{\sqrt{g}\psi^{4}}\biggl[\sigma^{ab}\sigma_{ab} - 
\frac{1}{6}(tr\pi)^{2}\psi^{12} - g\psi^{8}\biggl(R - 
8\frac{\grad^{2}\psi}{\psi}\biggr)\biggr] -
 2N_{a}\grad_{b}\pi^{ab} - \xi^{c}\grad_{c}tr\pi \eeq
As a consistency check let's find $\delg$ from this by varying with respect to 
$\pi^{ab}$. We get
\beq \delg = \frac{2N}{\sqrt{g}\psi^{4}}\biggl(\sigma_{ab} - 
\frac{1}{6}g_{ab}tr\pi\psi^{12}\biggr) + (KN)_{ab} + 
g_{ab}\grad_{c}\xi^{c} \eeq
as before. Thus, all is well. We may do all the usual variations here to get 
the constraints. Varying the Hamiltonian with respect to $g_{ab}$ gives us our 
evolution equation for $\pi^{ab}$. We get
\begin{equation}
\begin{split}
\frac{\partial\pi^{ab}}{\partial t} = & - 
N\sqrt{g}\psi^{4}\biggl(R^{ab} - g^{ab}\biggl(\rp\biggr)\biggr) \\ & - 
\frac{2N}{\sqrt{g}\psi^{4}}\biggl(\pi^{ac}\pi^{b}_{\;\;c} - \frac{1}{3}\pi^{ab}
tr\pi - \frac{1}{6}\pi^{ab}tr\pi\psi^{12}\biggr) \\ & + \sqrt{g}\psi
\biggl(\grad^{a}\grad^{b}(N\psi^{3}) - g^{ab}\grad^{2}(N\psi^{3})\biggr) \\ & 
+ N\sqrt{g}\psi^{3}\biggl(\grad^{a}\grad^{b}\psi + 3g^{ab}\grad^{2}\psi\biggr) 
 \\ & + 4g^{ab}\sqrt{g}\grad_{c}(N\psi^{3})\grad^{c}\psi - 
6\sqrt{g}\grad^{(a}(N\psi^{3})\grad^{b)}\psi
\\ & + \grad_{c}\biggl(\pi^{ab}N^{c}\biggr) - \pi^{bc}\grad_{c}N^{a} - 
\pi^{ac}\grad_{c}N^{b} \\ & - \biggl(\pi^{ab} - 
\frac{1}{2}g^{ab}tr\pi\biggr)\grad_{c}\xi^{c} 
\end{split}
 \eeq
We may use the evolution equations to find $\frac{\partial trp}{\partial t}$ 
quite easily. (Of course, we need the evolution equations to propagate all of 
the constraints. We will deal with the others later.) We find that
\beq \frac{\partial trp}{\partial t} = 0 \eeq
using the lapse-fixing equation. Thus we have that $trp =$ constant both 
spatially \emph{and} temporally!! We could proceed to check propagation of the 
constraints here but it will be easier and more instructive to do a little more
 work first.
\\ \\
Since $trp$ is identically a constant our dynamical data consists of $g_{ab}$ 
and $\sigma^{ab}$. Thus, it may prove useful to have an evolution equation for 
$\sigma^{ab}$ rather than the full $\pi^{ab}$. It is reasonably straightforward
 to do this. Firstly we note that
\beq \frac{\partial\sigma^{ab}}{\partial t} = 
\frac{\partial\pi^{ab}}{\partial t} - 
\frac{1}{3}\frac{\partial g^{ab}tr\pi}{\partial t} \eeq 
Working through the details gives us
\beq \begin{split} \frac{\partial\sigma^{ab}}{\partial t} = & - 
N\sqrt{g}\psi^{4}\biggl(R^{ab} - \frac{1}{3}g^{ab}\biggl(\rp\biggr)\biggr) - 
\frac{2N}{\sqrt{g}\psi^{4}}\sigma^{ac}\sigma^{b}_{\;\;c} \\ & + \sqrt{g}\psi
\biggl(\grad^{a}\grad^{b}(N\psi^{3}) - 
\frac{1}{3}g^{ab}\grad^{2}(N\psi^{3})\biggr) \\ & 
+ N\sqrt{g}\psi^{3}\biggl(\grad^{a}\grad^{b}\psi + 
\frac{7}{3}g^{ab}\grad^{2}\psi\biggr) 
 \\ & + 4g^{ab}\sqrt{g}\grad_{c}(N\psi^{3})\grad^{c}\psi - 
6\sqrt{g}\grad^{(a}(N\psi^{3})\grad^{b)}\psi 
\\ & + \grad_{c}\bigl(\sigma^{ab}N^{c}\biggr) -  \sigma^{bc}\grad_{c}N^{a} - 
 \sigma^{ac}\grad_{c}N^{b} \\ & - \sigma^{ab}\grad_{c}\xi^{c} + 
\frac{N\psi^{8}}{3\sqrt{g}}\sigma^{ab}tr\pi \end{split} \eeq
\section{Jacobi Action}
We can also find the Jacobi action of this theory. Recall the (3+1) Lagrangian,
\beq \mathcal{L} = N\sqrt{g}\psi^{4}\biggl(\rp + S^{ab}S_{ab} - 
\frac{2}{3}\psi^{-12}(trB)^{2}\biggr) \eeq
We can write this as
\beq \mathcal{L} = \sqrt{g}\psi^{4}\biggl[N\biggl(\rp\biggr) + 
\frac{1}{4N}\biggl(\Sigma^{ab}\Sigma_{ab} - 
\frac{2}{3}\psi^{-12}(tr\beta)^{2}\biggr)\biggr] \eeq
where $ \Sigma_{ab} = - 2NS_{ab}$ and $\beta_{ab} = -2NB_{ab}$. We now 
extremise with respect to $N$. This gives us,
\beq N = \underline{+}\frac{1}{2}\biggl(\Sigma^{ab}\Sigma_{ab} - 
\frac{2}{3}\psi^{-12}(tr\beta)^{2}\biggr)^{\frac{1}{2}}
\biggl(\rp\biggr)^{-\frac{1}{2}} \eeq
Substituting this back into the action gives us
\beq \label{nct} S = \underline{+}\int 
d\lambda\int\sqrt{g}\psi^{4}\sqrt{\rp}\sqrt{T}d^{3}x \eeq
where $ T = \biggl(\Sigma^{ab}\Sigma_{ab} - 
\frac{2}{3}\psi^{-12}(tr\beta)^{2}\biggr) $.
\\ \\
We can do all the usual variations here: $N^{a}$, $\xi^{c}$ and $\psi$. 
These give the momentum constraint, the conformal constraint and the 
lapse-fixing condition respectively. When we find the canonical momentum 
$\pi^{ab}$ we can ``square'' it to give the ``Hamiltonian constraint.''
\\ \\
So far, so good. We shall rarely use the Jacobi form of the action here but 
from a thin-sandwich point of view it is important and may well be of use in 
future work. Let's move on.
\section{Conformally Related Solutions}
We can do almost exactly the same thing here as we did in the section with the 
same name in Chapter 1. Suppose we start with initial data 
$\{g_{ab},\sigma^{ab},trp\}$ obeying the initial data conditions (\ref{c2new}) 
and (\ref{c3new}). We then solve (\ref{c1new}) for $\psi$
\\ \\
Suppose instead that we start with the conformally related initial data 
$\{h_{ab},\rho^{ab},trp\} = \{\alpha^{4}g_{ab},\alpha^{-4}\sigma^{ab},trp\}$. 
These automatically satisfy the initial data conditions by the conformal 
invariance. We now solve the Hamiltonian constraint for the conformal ``field''
 $\chi$, say. Just like before it can be shown that $\chi = 
\frac{\psi}{\alpha}$. Thus, yet again,
\beq \psi^{4}g_{ab} = \chi^{4}h_{ab} \eeq
and
\beq \psi^{-4}\sigma^{ab} = \chi^{-4}\rho^{ab} \eeq
Again we label these as $\wt{g_{ab}}$ and $\wt{\rho^{ab}}$ and put a hat over 
$trp$ also (for clarity). The constraints become
\beq \wt{\sigma_{ab}}\wt{\sigma^{ab}} - \frac{1}{6}\wt{(tr\pi)}^{2} - 
\wt{g}\wt{R} = 0 \eeq
\beq \wt{\grad_{b}}\wt{\pi^{ab}} = 0 \eeq
\beq \wt{\grad_{c}}\wt{tr\pi} = 0 \eeq
\beq \wt{N}\wt{R} - \wt{\grad^{2}}\wt{N} + \frac{(\wt{trp})^{2}}{4} = 0 \eeq
\\
Consider GR in the CMC gauge. The constraints are
\beq \sigma_{ab}\sigma^{ab} - \frac{1}{6}(tr\pi)^{2} - gR = 0 \eeq
\beq \grad_{b}\pi^{ab} = 0 \eeq
\beq \grad_{c}tr\pi = 0 \eeq
Evolution of the CMC condition gives
\beq NR - \grad^{2}N + \frac{(trp)^{2}}{4} = C \eeq
The similarities are \emph{quite} striking.
\subsection{What of $\xi^{c}$?}
Precious little has been revealed about what $\xi^{c}$ may be or even how it 
transforms. This needs to be addressed. First let's recall that we demanded 
that
\beq trB \gt \omega^{-8}trB \eeq
under a conformal transformation. This will be enough to reveal the 
transformation properties of $\xi^{c}$. Taking the trace gives us
\beq trB = - \frac{1}{2N}\biggl(g^{ab}\delg - g^{ab}(KN)_{ab} - 
3\grad_{c}\xi^{c}\biggr) \eeq
Under a conformal transformation we get
\beq 
\begin{split}  \omega^{-8}trB = & - \frac{1}{2\omega^{2}N}\biggl(g^{ab}\delg + 
12\frac{\dot{\omega}}{\omega} - \omega^{-4}g^{ab}\biggl(\omega^{4}(KN)_{ab} + 
4\omega^{3}\omega_{,c}N^{c}g_{ab}\biggr) - 3\bar{\grad_{c}}\bar{\xi^{c}}\biggr)
\\ = & - \frac{1}{2\omega^{2}N}\biggl(g^{ab}\delg - g^{ab}(KN)_{ab} - 
3\grad_{c}\xi^{c}\biggr) \\ & - \frac{3}{2\omega^{2}N}\grad_{c}\xi^{c} + 
\frac{3}{2\omega^{2}N}\bar{\grad_{c}}\bar{\xi^{c}} - 
\frac{6}{\omega^{3}N}\biggl(\dot{\omega} - \omega_{,c}N^{c}\biggr) \\
= & \omega^{-2}trB + \frac{3}{2\omega^{2}N}\biggl(\bar{\grad_{c}}\bar{\xi^{c}}
 - 3\grad_{c}\xi^{c} - 
\frac{4}{\omega}\biggl(\dot{\omega} - \omega_{,c}N^{c}\biggr)\biggr) 
\end{split} \eeq
Thus,
\beq \frac{3}{2\omega^{2}N}\biggl(\bar{\grad_{c}}\bar{\xi^{c}}
 - 3\grad_{c}\xi^{c} - \frac{4}{\omega}
\biggl(\dot{\omega} - \omega_{,c}N^{c}\biggr)\biggr)
= - \frac{1}{\omega^{2}N}trB\biggl(1 - \omega^{-6}\biggr) \eeq
and so
\beq \label{txic} \bar{\grad_{c}}\bar{\xi^{c}} = \grad_{c}\xi^{c} + 
\frac{4}{\omega}\biggl(\dot{\omega} - \omega_{,c}N^{c}\biggr) - 
\frac{2N}{3}trB\biggl(1 - \omega^{-6}\biggr) \eeq
This tells us how things transform but not what $\xi^{c}$ itself actually is. 
We \emph{can} find this though.
\\ \\
Let's write the evolution equations in the physical representation. It can be 
verified that they are
\beq \frac{\partial \wt{g_{ab}}}{\partial t} = \frac{2\wt{N}}{\sqrt{\wt{g}}}
\biggl(\wt{\sigma_{ab}} - \frac{1}{6}\wt{g_{ab}}\wt{tr\pi}\biggr) + 
\wt{(KN)_{ab}} + \wt{g_{ab}}\wt{\grad_{c}}\wt{\xi^{c}} \eeq
and
\beq
\begin{split} \frac{\partial\wt{\sigma^{ab}}}{\partial t} = & - 
\wt{N}\sqrt{\wt{g}}\biggl(\wt{R^{ab}} - \frac{1}{3}\wt{g^{ab}}\wt{R}\biggr) - 
\frac{2\wt{N}}{\sqrt{\wt{g}}}
\wt{\sigma^{ac}}\wt{\sigma^{b}_{\;\;c}} \\ & + \sqrt{\wt{g}}
\biggl(\wt{\grad^{a}}\wt{\grad^{b}}\wt{N} - 
\frac{1}{3}\wt{g^{ab}}\wt{\grad^{2}}\wt{N}
\biggr) \\ & + \wt{\grad_{c}}\biggl(\wt{\sigma^{ab}}\wt{N^{c}}\biggr) - 
\wt{\sigma^{bc}}\wt{\grad_{c}}\wt{N^{a}} - 
\wt{\sigma^{ac}}\wt{\grad_{c}}\wt{N^{b}} 
\\ & + \frac{\wt{N}}{3\sqrt{\wt{g}}}\wt{\sigma^{ab}}\wt{tr\pi} - 
\wt{\sigma^{ab}}\wt{\grad_{c}}\wt{\xi^{c}}\end{split} \eeq
We require the evolution equations to propagate the constraints. However, when
 we check this it turns out that we are \emph{forced} to set 
$\wt{\grad_{c}}\wt{\xi^{c}}$ to zero. However, this means that we have
\beq \grad_{c}\xi^{c} + 
\frac{4}{\omega}(\dot{\psi} - \psi_{,c}N^{c}) - \frac{2N}{3}trB(1 - \psi^{-6}) 
= 0 \eeq
by (\ref{txic}). Thus we have
\beq \grad_{c}\xi^{c} =  - \frac{4}{\psi}(\dot{\psi} - \psi_{,c}N^{c}) + 
\frac{2N}{3}trB(1 - \psi^{-6}) \eeq
That is,
\beq \grad_{c}\xi^{c} = \theta + \frac{2N}{3}trB(1 - \psi^{-6}) \eeq
where $\theta$ is as in the original theory. Thus, the exact form of $\xi^{c}$ 
is determined. We needed $\grad_{c}\xi^{c}$ to be zero in the physical 
representation for constraint propagation and so we should check that this is 
the case with our newly found expression for $\grad_{c}\xi^{c}$. We can check 
this easily. In the physical representation $\theta = 0$ and $\psi = 1$. Thus, 
we do have that $\wt{\grad_{c}}\wt{\xi^{c}}$ is zero.
\\ \\
It is vital to note that this is strictly a \emph{POST-VARIATION} 
identification. If we use this form for $\xi^{c}$ in the action we will run 
into problems, not least an infinite sequence in the variation of $trB$ with 
respect to $\xi^{c}$. (This is because we would have $trB$ defined in terms of 
$trB$ itself.) We see that $\xi^{c}$ is intimately related with how $\psi$ 
changes from slice to slice.
\\ \\
Our constraints in the physical representation are
\beq \sigma_{ab}\sigma^{ab} - \frac{1}{6}(tr\pi)^{2} - gR = 0 \eeq
\beq \grad_{b}\pi^{ab} = 0 \eeq
\beq \grad_{c}tr\pi = 0 \eeq
\beq NR - \grad^{2}N + \frac{N(trp)^{2}}{4} = 0 \eeq
and our evolution equations are
\beq \frac{\partial g_{ab}}{\partial t} = \frac{2N}{\sqrt{g}}
\biggl(\sigma_{ab} - \frac{1}{6}g_{ab}tr\pi\biggr) + (KN)_{ab} \eeq
\beq \begin{split} \frac{\partial\sigma^{ab}}{\partial t} = & - 
N\sqrt{g}\biggl(R^{ab} - \frac{1}{3}g^{ab}R\biggr) - 
\frac{2N}{\sqrt{g}}\sigma^{ac}\sigma^{b}_{\;\;c} \\ & + \sqrt{g}
\biggl(\grad^{a}\grad^{b}N - \frac{1}{3}g^{ab}\grad^{2}N
\biggr) \\ & + \grad_{c}(\sigma^{ab}N^{c}) - 
\sigma^{bc}\grad_{c}N^{a} - 
\sigma^{ac}\grad_{c}N^{b} 
\\ & + \frac{N}{3\sqrt{g}}\sigma^{ab}tr\pi \end{split} \eeq
(The hats are removed for simplicity.) These are \emph{identical} to those in 
GR in the CMC gauge (with $trp$ a temporal constant).
\section{Topological Considerations}
Mimicking the section in Chapter one we find in the same way that if the 
manifold is compact without boundary we get frozen dynamics. In the 
asymptotically flat case we have no such problem and this will prove to be 
important in solar system tests of the theory. In the problematic case we can 
resolve the issue in the same manner as before although, it is a \emph{little} 
more complicated this time.
\subsection{Integral Inconsistencies (Slight Return)}
The root of the integral inconsistency is in the lapse-fixing equation. If we 
integrate this equation we find that the only solution is $N \equiv 0$. That 
is, we have frozen dynamics. The resolution to this in regular CG was to 
introduce a volume term in the denominator of the Lagrangian. Actually, the key
 is to keep the Lagrangian homogeneous in $\psi$ using different powers of the 
volume. The volume of a hypersurface here is given by
\beq V = \int \sqrt{g}\psi^{6}\;d^{3}x \eeq
In the original theory the Lagrangian has an overall factor of $\psi^{4}$ and 
so we need to divide by $V^{2/3}$ to keep homogeneity in $\psi$. There is no 
such overall factor in the new theory and so it is not as straightforward. The 
key is to treat the two parts of the Lagrangian separately. We try
\beq \mathcal{L}_{1} = \frac{N\sqrt{g}\psi^{4}}{V^{n}}\biggl(R - 
8\frac{\grad^{2}\psi}{\psi} + S^{ab}S_{ab}\biggr) \eeq
and
\beq \mathcal{L}_{2} = -\frac{2}{3}\frac{N\sqrt{g}\psi^{-8}}{V^{m}}(trB)^{2} 
\eeq
and we determine $n$ and $m$ from the homogeneity requirement. Thus we have 
that $n = \frac{2}{3}$ and $m = -\frac{4}{3}$. Using this result our Lagrangian
is now
\beq \label{lag3} \mathcal{L} = \frac{N\sqrt{g}\psi^{4}}{V^{2/3}}\biggl(R - 
8\frac{\grad^{2}\psi}{\psi} + S^{ab}S_{ab} - 
\frac{2}{3}\psi^{-12}(trB)^{2}V^{2}\biggr) \eeq
\subsection{New Constraints}
We go about things in exactly the same manner as before. The momentum is found 
to be
\beq \pi^{ab} = - \frac{\sqrt{g}\psi^{4}}{V^{2/3}}S^{ab} + 
\frac{2}{3}\sqrt{g}\psi^{-8}V^{4/3}g^{ab}trB \eeq
The constraints are (almost) unchanged. They are
\beq \sigma_{ab}\sigma^{ab} - \frac{\psi^{12}(tr\pi)^{2}}{6V^{2}} - 
\frac{g\psi^{8}}{V^{4/3}}\biggl(\rp\biggr) = 0 \eeq
\beq \grad_{b}\pi^{ab} = 0 \eeq
\beq \grad_{c}tr\pi = 0 \eeq
The lapse-fixing equation is
\beq \frac{N\sqrt{g}\psi^{3}}{V^{2/3}}\biggl(R - 7\frac{\grad^{2}\psi}{\psi}
\biggr) - \sqrt{g}\frac{\grad^{2}(N\psi^{3})}{V^{2/3}} - 
\sqrt{g}\frac{C\psi^{5}}{2V^{2/3}} + 
\sqrt{g}N\psi^{-9}(trB)^{2}V^{4/3} - 
\frac{2}{3}\sqrt{g}D\psi^{5}V^{4/3} = 0 \eeq
where
\beq C = \int \frac{N\sqrt{g}\psi^{4}}{V}\biggl(R - 
8\frac{\grad^{2}\psi}{\psi} + S^{ab}S_{ab}\biggr) \; d^{3}x \eeq
and
\beq D = \int \frac{N\sqrt{g}\psi^{-8}}{V}(trB)^{2}\;d^{3}x \eeq
The $C$ and $D$ terms result from the variations of the volume. Rearranging the
 lapse-fixing equation we get
\beq \frac{N\sqrt{g}\psi^{3}}{V^{2/3}}\biggl(R - 7\frac{\grad^{2}\psi}{\psi}
\biggr) - \sqrt{g}\frac{\grad^{2}(N\psi^{3})}{V^{2/3}} 
\sqrt{g}N\psi^{-9}(trB)^{2}V^{4/3} = 
\frac{\sqrt{g}\psi^{5}}{2V^{2/3}}\biggl(C + \frac{4}{3}DV^{2}\biggr) \eeq
Integrating across this expression gives \emph{no} problem. The inconsistency 
has been removed.
\section{The Hamiltonian Formulation}
We should consider the evolution equations again now that we have changed the 
action. First of all the momentum is now given by
\beq \pi^{ab} = -\frac{\sqrt{g}\psi^{4}}{V^{2/3}}S^{ab} + 
\frac{2}{3}\sqrt{g}\psi^{-8}g^{ab} trB V^{4/3} \eeq
The new Hamiltonian is
\beq \mathcal{H} = 
\frac{NV^{2/3}}{\sqrt{g}\psi^{4}}\biggl[\sigma^{ab}\sigma_{ab} - 
\frac{(tr\pi)^{2}\psi^{12}}{6V^{2}} - \frac{g\psi^{8}}{V^{4/3}}\biggl(R - 
8\frac{\grad^{2}\psi}{\psi}\biggr)\biggr] -
 2N_{a}\grad_{b}\pi^{ab} - \xi^{c}\grad_{c}tr\pi \eeq
The evolution equations are then
\beq \delg = \frac{2NV^{2/3}}{\sqrt{g}\psi^{4}}\biggl(\sigma_{ab} - 
\frac{g_{ab}tr\pi\psi^{12}}{6V^{2}}\biggr) + (KN)_{ab} + 
g_{ab}\grad_{c}\xi^{c} \eeq
and
\beq
\begin{split} \frac{\partial\pi^{ab}}{\partial t} = & - 
\frac{N\sqrt{g}\psi^{4}}{V^{2/3}}\biggl(R^{ab} - 
g^{ab}\biggl(\rp\biggr)\biggr) \\ & - 
\frac{2NV^{2/3}}{\sqrt{g}\psi^{4}}\biggl(\pi^{ac}\pi^{b}_{\;\;c} - 
\frac{1}{3}\pi^{ab}tr\pi - \frac{\pi^{ab}tr\pi\psi^{12}}{6V^{2}}\biggr) \\ & + 
\frac{\sqrt{g}\psi}{V^{2/3}}\biggl(\grad^{a}\grad^{b}(N\psi^{3}) - 
g^{ab}\grad^{2}(N\psi^{3})\biggr) \\ & + 
\frac{N\psi^{3}\sqrt{g}}{V^{2/3}}\biggl(\grad^{a}\grad^{b}\psi + 
3g^{ab}\grad^{2}\psi\biggr)
\\ & 
+ 4\frac{\sqrt{g}}{V^{2/3}}g^{ab}\grad_{c}(N\psi^{3})\grad^{c}\psi - 
6\frac{\sqrt{g}}{V^{2/3}}\grad^{(a}(N\psi^{3})\grad^{b)}\psi 
\\ & + \grad_{c}\biggl(\pi^{ab}N^{c}\biggr) -  \pi^{bc}\grad_{c}N^{a} -  
\pi^{ac}\grad_{c}N^{b} \\ & - (\pi^{ab} - 
\frac{1}{2}g^{ab}tr\pi)\grad_{c}\xi^{c} - 
\frac{2}{3}\frac{\sqrt{g}\psi^{6}g^{ab}}{V^{2/3}}C \end{split} \eeq
where
\beq C = \biggl<N\sqrt{g}\psi^{4}\biggl(\rp + 
\frac{\psi^{4}(trp)^{2}}{4V^{2/3}}\biggr)\biggr> \eeq
\\ \\
We can again take the time derivative of $trp$ and find yet again that
\beq \frac{\partial trp}{\partial t} = 0 \eeq
Thus, our dynamic data will once again be $\{g_{ab},\sigma^{ab}\}$ and so we 
want to find the evolution equation for $\sigma^{ab}$ again. Slogging through 
we get
\beq
\begin{split} \frac{\partial\sigma^{ab}}{\partial t} = & - 
\frac{N\sqrt{g}\psi^{4}}{V^{2/3}}\biggl(R^{ab} - 
\frac{1}{3}g^{ab}\biggl(\rp\biggr)\biggr) - \frac{2NV^{2/3}}{\sqrt{g}
\psi^{4}}\sigma^{ac}\sigma^{b}_{\;\;c} \\ & + \frac{\sqrt{g}\psi}{V^{2/3}}
\biggl(\grad^{a}\grad^{b}(N\psi^{3}) - 
\frac{1}{3}g^{ab}\grad^{2}(N\psi^{3})\biggr) \\ & + 
\frac{N\sqrt{g}\psi^{3}}{V^{2/3}}\biggl(\grad^{a}\grad^{b}\psi + 
\frac{7}{3}g^{ab}\grad^{2}\psi\biggr)
\\ & + \frac{4\sqrt{g}}{V^{2/3}}g^{ab}\grad_{c}(N\psi^{3})\grad^{c}\psi - 
\frac{6\sqrt{g}}{V^{2/3}}\grad^{(a}(N\psi^{3})\grad^{b)}\psi 
\\ & + \grad_{c}\biggl(\sigma^{ab}N^{c}\biggr) -  \sigma^{bc}\grad_{c}N^{a} - 
 \sigma^{ac}\grad_{c}N^{b} \\ & - \sigma^{ab}\grad_{c}\xi^{c} + 
\frac{N\psi^{8}}{3\sqrt{g}V^{4/3}}\sigma^{ab}tr\pi \end{split} \eeq
Note that the term with $C$ has dropped out.
\\ \\
The physical representation is achieved either by the naive substitution of 
$\psi = 1$ and $\grad_{c}\xi^{c} = 0$ or by doing it the longer more correct 
way. The result is the same in either case. The momentum is
\beq \pi^{ab} = - \frac{\sqrt{g}}{V^{2/3}}S^{ab} + 
\frac{2}{3}\sqrt{g}g^{ab}trKV^{4/3} \eeq
Thus
\beq \label{oops} \sigma^{ab} = -\frac{\sqrt{g}}{V^{2/3}}S^{ab} \;\; \text{and}
 \;\; tr\pi = 2\sqrt{g}(trK)V^{4/3} \eeq
The constraints are
\beq \sigma^{ab}\sigma_{ab} - \frac{1}{6}\frac{(tr\pi)^{2}}{V^{2}} = 
\frac{gR}{V^{4/3}} \eeq
\beq \grad_{b}\pi^{ab} = 0 \eeq
\beq \label{cc2} \grad_{c}trp = 0 \eeq
\beq NR -\grad^{2}N + \frac{N(trp)^{2}}{4V^{2/3}} =  C \eeq
where we now have 
$C = \biggl<N\biggl(R + \frac{(trp)^{2}}{4V^{2/3}}\biggr)\biggr>$. The 
evolution equations are
\beq \delg = \frac{2NV^{2/3}}{\sqrt{g}}\biggl(\sigma_{ab} - 
\frac{g_{ab}tr\pi}{6V^{2}}\biggr) + (KN)_{ab} \eeq
and
\beq
\begin{split} \frac{\partial\sigma^{ab}}{\partial t} = & - 
\frac{N\sqrt{g}}{V^{2/3}}\biggl(R^{ab} - 
\frac{1}{3}g^{ab}R\biggr) - \frac{2NV^{2/3}}{\sqrt{g}
}\sigma^{ac}\sigma^{b}_{\;\;c} \\ & + \frac{\sqrt{g}\psi}{V^{2/3}}
\biggl(\grad^{a}\grad^{b}N - \frac{1}{3}g^{ab}\grad^{2}N\biggr) \\ & + 
\grad_{c}\biggl(\sigma^{ab}N^{c}\biggr) - \sigma^{bc}\grad_{c}N^{a} - 
\sigma^{ac}\grad_{c}N^{b} \\ & + \frac{N}{3\sqrt{g}V^{4/3}}\sigma^{ab}tr\pi 
\end{split} \eeq
\section{The Volume}
This theory was inspired by the need to recover expansion. After all this work,
 have we succeeded? The time derivative of the volume is
\beq \begin{split} \frac{\partial V}{\partial t} = & 
\int\frac{1}{2}\sqrt{g}g^{ab}\delg\;\;d^{3}x \\ = & - 
\int\frac{1}{2}\frac{N\sqrt{g}trp}{V^{4/3}}\;\;d^{3}x \\ = & - 
\frac{trp\Big<N\Big>}{2V^{1/3}} \end{split} \eeq
Thus, we have recovered expansion. The big test of the compact without boundary
 theory will be to study the cosmological solutions and this will be the focus 
of a later chapter.
\section{Jacobi Action}
For completeness let's find the Jacobi action for the compact theory. Without 
going through each step let's simply require homogeneity in $\psi$. The Jacobi 
action for the non-compact theory was (\ref{nct})
\beq S = \underline{+}\int d\lambda\int\sqrt{g}\psi^{4}\sqrt{\rp}\sqrt{T}d^{3}x
\eeq
where $ T = \biggl(\Sigma^{ab}\Sigma_{ab} - 
\frac{2}{3}\psi^{-12}(tr\beta)^{2}\biggr)$ where $\Sigma^{1b} = - 2NS^{ab}$ and
 $\beta^{ab} = - 2NB^{ab}$. Applying the homogeneity requirement gives
\beq S = \underline{+}\int d\lambda\int
\frac{\sqrt{g}\psi^{4}\sqrt{\rp}\sqrt{T}d^{3}x}{V^{2/3}} \eeq
where $ T = \biggl(\Sigma^{ab}\Sigma_{ab} - 
\frac{2}{3}\psi^{-12}(tr\beta)^{2}V^{2}\biggr) $.
\\ \\
Everything else emerges as before.
\section{Comparison with GR}
In the earlier ``static'' conformal theory we saw that the labelling
\beq \widehat{\pi^{ab}} = V^{2/3}\pi^{ab} \eeq
made the theory appear incredibly similar to GR. A similar labelling is 
possible here. Define
\beq \widehat{\sigma^{ab}} = V^{2/3}\sigma^{ab} \eeq
and
\beq \widehat{tr\pi} = \frac{tr\pi}{V^{1/3}} \eeq
With this rebelling the constraints are
\beq \widehat{\sigma^{ab}}\widehat{\sigma_{ab}} - 
\frac{1}{6}(\widehat{tr\pi})^{2} = gR \eeq
\beq \grad_{b}\widehat{\pi^{ab}} = 0 \eeq
\beq \grad_{c}\widehat{trp} = 0 \eeq
and the lapse-fixing equation is
\beq NR - \grad^{2}N + \frac{N(\widehat{trp})^{2}}{4} = C \eeq
where $C = \biggl<N\biggl(R + \frac{(\widehat{trp})^{2}}{4}\biggr)\biggr>$. 
These are \emph{identical} to GR in the CMC gauge. The evolution equations are
\beq \delg = \frac{2N}{\sqrt{g}}\biggl(\widehat{\sigma_{ab}} - 
\frac{g_{ab}\widehat{tr\pi}}{6V}\biggr) + (KN)_{ab} \eeq
and
\beq
\begin{split} \frac{\partial\widehat{\sigma^{ab}}}{\partial t} = & 
V^{2/3}\frac{\partial\sigma^{ab}}{\partial t} + 
\frac{2}{3V^{1/3}}\frac{\partial V}{\partial t} \\ = & - 
N\sqrt{g}\biggl(R^{ab} - \frac{1}{3}g^{ab}R\biggr) - 
\frac{2N}{\sqrt{g}}\widehat{\sigma^{ac}}\widehat{\sigma^{b}_{\;\;c}} \\ & + 
\sqrt{g}\psi\biggl(\grad^{a}\grad^{b}N - \frac{1}{3}g^{ab}\grad^{2}N\biggr) \\ 
& + \grad_{c}(\widehat{\sigma^{ab}}N^{c}) - 
\widehat{\sigma^{bc}}\grad_{c}N^{a} - 
\widehat{\sigma^{ac}}\grad_{c}N^{b} \\ & + 
\frac{\biggl(N - \Big<N\Big>\biggr)}{3\sqrt{g}V}
\widehat{\sigma^{ab}}\widehat{tr\pi} \end{split} \eeq
In GR these are
\beq \delg = \frac{2N}{\sqrt{g}}\biggl(\sigma_{ab} - 
\frac{g_{ab}tr\pi}{6}\biggr) + (KN)_{ab} \eeq
and
\beq
\begin{split} \frac{\partial\sigma^{ab}}{\partial t} = & - 
N\sqrt{g}\biggl(R^{ab} - \frac{1}{3}g^{ab}R\biggr) - 
\frac{2N}{\sqrt{g}}\sigma^{ac}\sigma^{b}_{\;\;c} \\ & + 
\sqrt{g}\psi\biggl(\grad^{a}\grad^{b}N - \frac{1}{3}g^{ab}\grad^{2}N\biggr) \\ 
& + \grad_{c}(\sigma^{ab}N^{c}) - 
\sigma^{bc}\grad_{c}N^{a} - 
\sigma^{ac}\grad_{c}N^{b} \\ & + 
\frac{N}{3\sqrt{g}}\sigma^{ab}tr\pi \end{split} \eeq
There are very few differences between the theories. We shall compare the two 
theories later in purely geometric terms, that is, in terms of the metric and 
curvature rather than the momentum. Let us leave this for now.
\section{Time}
Recall the York time (\ref{yortim})
\beq \tau = \frac{2}{3}trp \eeq
In GR this was a good notion of time. However in this theory we have that $trp$
 is identically constant. Thus it cannot be used as a notion of time. We note 
now though that unlike in GR, for us the volume is monotonically increasing. 
(In fact, it goes from $0$ to $\infty$ as we shall see when discussing 
cosmology later.) Of course, the volume \emph{must} be constant on any 
hypersurface and so the volume is a good notion of time in this theory. This 
may be extremely beneficial in a quantisation program.
\section{Light Cones}
So far the theory is quite promising. There are a number of things that must 
carry over from GR though if it is to be taken seriously. One of these is that 
the speed of propagation of the wave front must be unity (the speed of light). 
The easiest way to check this is to consider the evolution equations. Let's 
consider the case in GR briefly. The corresponding case in the conformal theory
 will work in almost exactly the same way.
\\ \\
The evolution equation for $g_{ab}$ in GR is
\beq \delg = \frac{2N}{\sqrt{g}}\biggl(\pi_{ab} - 
\frac{1}{2}g_{ab}tr\pi\biggr) + (KN)_{ab} \eeq
Inverting this we find that
\beq \pi_{ab} = 
\frac{\sqrt{g}}{2N}\frac{\partial g_{ab}}{\partial t} \eeq
We will be working here to leading order in the derivatives which is the reason
 for only omitting the other terms. Differentiating both sides gives
\beq \frac{\partial\pi_{ab}}{\partial t} = \frac{\sqrt{g}}{2N}
\frac{\partial^{2}g_{ab}}{\partial t^{2}} \eeq
Now substituting this into the evolution equation for $\pi_{ab}$ gives us
\beq \label{sol} \frac{\sqrt{g}}{2N}\frac{\partial^{2}g_{ab}}{\partial t^{2}} =
 - N\sqrt{g}\biggl(R_{ab} - \frac{1}{2}g_{ab}R\biggr) \eeq
(Note: The alternate form of the evolution equation is used here with the 
factor of $\frac{1}{2}$ on $R$.)
\\ \\
Now,
\beq \biggl(R_{ab} - \frac{1}{2}g_{ab}R\biggr) = \frac{1}{2}g^{cd}
\biggl[g_{bd,ac} + g_{ac,bd} - g_{ab,cd} - g_{cd,ab} - g_{ab}g^{ef}
\biggl(g_{ec,fd} - g_{ef,cd}\biggr)\biggr] \eeq
again only using leading order in the derivatives. We are concerned with the 
transverse traceless part of $g_{ab}$ which we'll label as $g_{ab}^{TT}$. The 
only relevant part is then
\beq - \frac{1}{2}g^{cd}g^{TT}_{ab,cd} \eeq
which we'll write as
\beq -\frac{1}{2}\frac{\partial^{2}g^{TT}_{ab}}{\partial x^{2}} \eeq
All the other terms are cancelled either through the transverse or traceless 
properties. Using only the $TT$ part in the time derivatives also gives us
\beq \frac{1}{2N^{2}}\frac{\partial^{2}g^{TT}_{ab}}{\partial t^{2}} = 
\frac{1}{2}\frac{\partial^{2}g^{TT}_{ab}}{\partial x^{2}} \eeq
This is a wave equation with wave speed $1$. Thus we get gravitational 
radiation! Various details are omitted here but the essence of the idea is 
quite clear. Let's consider the conformal theory. We'll use the compact without
 boundary theory (that is, the one with the volume terms).
\\ \\
The evolution equation for $g_{ab}$ can be inverted to get
\beq \sigma_{ab}\frac{\sqrt{g}}{2NV^{2/3}}\delg + ... \eeq
Differentiating both sides gives
\beq \frac{\partial\sigma_{ab}}{\partial t} = \frac{\sqrt{g}}{2NV^{2/3}}
\frac{\partial^{2}g_{ab}}{\partial t^{2}} \eeq
again, working only to leading order in the derivatives. Substituting this into
 the evolution equation for $\sigma_{ab}$ gives us
\beq \frac{\sqrt{g}}{2NV^{2/3}}
\frac{\partial^{2}g_{ab}}{\partial t^{2}} = 
\frac{N\sqrt{g}}{V^{2/3}}\biggl(R_{ab} - \frac{1}{2}g_{ab}R\biggr) \eeq
The volume terms cancel and we are left with the same equation as (\ref{sol}) 
above. In exactly the same way this becomes
\beq \frac{1}{N^{2}}\frac{\partial^{2}g^{TT}_{ab}}{\partial t^{2}} = 
\frac{\partial^{2}g^{TT}_{ab}}{\partial x^{2}} \eeq
Yet again, we have found a wave equation with speed $1$. Thus we have recovered
 gravitational radiation with wavefronts propagating at the speed of light. All
 is still well.
\section{Matter in General Relativity}
While the aim is to couple matter to gravity in the new theory it will be 
instructive to warm up by reviewing the corresponding cases in GR. We'll treat 
various different sources, namely, a cosmological constant, electromagnetism 
and dust dealing with each one in turn.
\subsection{Cosmological Constant}
This is the easiest of all matter sources. Let's take the Lagrangian for vacuum
GR to be
\beq \mathcal{L} = N\sqrt{g}\biggl(R + K^{ab}K_{ab} - (trK)^{2} \biggr) \eeq
The Lagrangian for a cosmological constant is simply
\beq \mathcal{L}_{cc} = - N\sqrt{g}\Lambda \eeq
and our full Lagrangian is
\beq \mathcal{L} = N\sqrt{g}\biggl(R - \Lambda + K^{ab}K_{ab} - 
(trK)^{2}\biggr) \eeq
There is no change to the momentum constraint. The Hamiltonian constraint 
changes in an easy way becoming
\beq K^{ab}K_{ab} - (trK)^{2} - \biggl(R - \Lambda\biggr) = 0 \eeq
In terms of the momentum this is
\beq \pi^{ab}\pi_{ab} - \frac{1}{2}(tr\pi)^{2} -g\biggl(R - \Lambda\biggr) = 0 
\eeq
The momentum constraint is unchanged ($\Lambda$ has no conjugate momentum). In 
fact, to see the changes here one simply substitutes $R - \Lambda$ wherever 
there was $R$. Thus there is no change to the evolution equation for $g_{ab}$ 
and there is only one simple change to the evolution equation for $\pi^{ab}$. 
Let's move on to electromagnetism.
\subsection{Electromagnetism}
The Lagrangian for electromagnetism is
\beq \mathcal{L}_{em} = N\sqrt{g}\biggl(U + T\biggr) \eeq
where
\beq U = -\frac{1}{4}\biggl(\grad_{b}A_{a} - 
\grad_{a}A_{b}\biggr)\grad^{b}A^{a} \eeq
and
\beq T = \frac{1}{4N^{2}}g^{ab}\biggl(\frac{\partial A_{a}}{\partial t} - 
\pounds_{\underline{N}}A_{a} - \grad_{a}\Phi\biggr)
\biggl(\frac{\partial A_{b}}{\partial t} - \pounds_{\underline{N}}A_{b} - 
\grad_{b}\Phi\biggr) \eeq
The full Lagrangian is then
\beq \mathcal{L} = N\sqrt{g}\biggl(R + U + K^{ab}K_{ab} - (trK)^{2} + 
T\biggr) \eeq
The Hamiltonian constraint is
\beq \pi^{ab}\pi_{ab} - \frac{1}{2}(tr\pi)^{2} + \pi^{c}\pi_{c} - 
g\biggl(R + U\biggr) = 0 \eeq
where $\pi^{c}$ is the momentum conjugate to $A_{c}$ given by
\beq \pi^{c} = 
\frac{\sqrt{g}}{2N}g^{ac}\biggl(\frac{\partial A_{a}}{\partial t} - 
\pounds_{\underline{N}} - \grad_{a}\Phi\biggr) \eeq
In more familiar language this becomes
\beq \pi^{ab}\pi_{ab} - \frac{1}{2}(tr\pi)^{2} -g\biggl(R - 
16\pi\rho_{r}\biggr) = 0 \eeq
where $\rho_{r}$ is the energy density of the radiation.
\\ \\
The variation with respect to $\Phi$ gives us the electromagnetic Gauss 
constraint
\beq \grad_{c}\pi^{c} = 0 \eeq
and using this the momentum constraint is unchanged
\beq \grad_{b}\pi^{ab} = 0 \eeq
\subsection{Dust}
The Lagrangian for dust is given by
\beq \mathcal{L}_{d} = \sqrt{-^{(4)}g}\rho_{d}
\biggl(g^{\alpha\beta}U_{\alpha}U_{\beta} + 1\biggr) \eeq
where $\rho_{d}$ is the rest mass density and $U_{\alpha}$ is the four velocity
 of the dust. This is written here in $4$-D form to show some of the 
properties. In particular the constraint which arises on variation with respect
 to $M$ is
\beq g^{\alpha\beta}U_{\alpha}U_{\beta} + 1 = 0 \eeq
Keeping this constraint will be important when we attempt to couple dust to the
 conformal theory. In any case the full Lagrangian is
\beq \mathcal{L} = N\sqrt{g}\biggl(R + K^{ab}K_{ab} - (trK)^{2} + 
\rho_{d}\biggl(g^{\alpha\beta}U_{\alpha}U_{\beta} + 1\biggr)\biggr) \eeq
The Hamiltonian constraint is
\beq \pi^{ab}\pi_{ab} - \frac{1}{2}(tr\pi)^{2} -g\biggl(R - \rho_{d}\biggr) = 0
 \eeq
The momentum constraint is unchanged.
\section{Matter and Conformal Gravity}
To approach this problem we'll proceed as in the GR cases above by simply 
considering each type of matter in turn. The changes are very straightforward 
in any case. We'll consider the case without the volume terms for simplicity. 
Inserting the volume terms will be easy by just requiring homogeneity in 
$\psi$.
\subsection{Cosmological Constant} 
The easiest thing to do is to take the GR Lagrangian and conformalise it. In GR
we had
\beq \mathcal{L} = N\sqrt{g}\biggl(R - \Lambda + K^{ab}K_{ab} - 
(trK)^{2}\biggr) \eeq
This becomes
\beq \mathcal{L} = N\psi^{2}\sqrt{g}\psi^{6}\biggl(\psi^{-4}\biggl(\rp\biggr) -
 \Lambda + \psi^{-4}S^{ab}S_{ab} - \psi^{-16}(trB)^{2}\biggr) \eeq
Simplifying this gives
\beq \mathcal{L} = N\sqrt{g}\psi^{4}\biggl(\rp - \psi^{4}\Lambda + S^{ab}S_{ab}
 - \psi^{-12}(trB)^{2}\biggr) \eeq
The Hamiltonian constraint becomes
\beq \sigma^{ab}\sigma_{ab} - \frac{1}{6}\psi^{12}(trp)^{2} - 
g\psi^{8}\biggl(\rp - \psi^{-4}\Lambda\biggr) = 0 \eeq
If we want the compact without boundary case we simply substitute in volumes so
as to achieve homogeneity in $\psi$. We get
\beq \sigma^{ab}\sigma_{ab} - \frac{1}{6}\frac{\psi^{12}(trp)^{2}}{V^{2}} - 
\frac{g\psi^{8}}{V^{4/3}}\biggl(\rp - \frac{\psi^{-4}\Lambda}{V^{2/3}}\biggr) =
 0 \eeq
In the physical representation this is
\beq \sigma^{ab}\sigma_{ab} - \frac{1}{6}\frac{(trp)^{2}}{V^{2}} - 
\frac{g}{V^{4/3}}\biggl(R - \frac{\Lambda}{V^{2/3}}\biggr) = 0 \eeq
So far so good.
\subsection{Electromagnetism}
The Lagrangian in GR was
\beq \mathcal{L} = N\sqrt{g}\biggl(R + U + K^{ab}K_{ab} - (trK)^{2} + 
T\biggr) \eeq
Conformalising this gives
\beq \mathcal{L} = N\psi^{2}\sqrt{g}\psi^{6}\biggl(\psi^{-4}\biggl(\rp\biggr) +
 \psi^{-8}U + \psi^{-4}S^{ab}S_{ab} - \psi^{-16}(trB)^{2} + \psi^{-8}T\biggr) 
\eeq
where $U$ and $T$ are unchanged from the GR case. Simplifying this gives us
\beq \mathcal{L} = N\sqrt{g}\psi^{4}\biggl(\rp + \psi^{-4}U + S^{ab}S_{ab} - 
\psi^{-12}(trB)^{2} + \psi^{-4}T\biggr) \eeq
The Hamiltonian constraint is
\beq \sigma^{ab}\sigma_{ab} - \frac{1}{6}\psi^{12}(trp)^{2} + 
\psi^{4}\pi^{c}\pi_{c} - g\psi^{8}\biggl(\rp + \psi^{-4}U\biggr) = 0 \eeq
where $\pi^{c}$ is the momentum conjugate to $A_{c}$ as before. In the compact 
without boundary case we get
\beq \sigma^{ab}\sigma_{ab} - \frac{1}{6}\frac{\psi^{12}(trp)^{2}}{V^{2}} + 
\frac{\psi^{4}\pi^{c}\pi_{c}}{V^{2/3}} - 
\frac{g\psi^{8}}{V^{4/3}}\biggl(\rp + \psi^{-4}UV^{2/3}\biggr) = 0 \eeq
In the physical representation this is
\beq \sigma^{ab}\sigma_{ab} - \frac{1}{6}\frac{(trp)^{2}}{V^{2}} + 
\frac{\pi^{c}\pi_{c}}{V^{2/3}} - 
\frac{g}{V^{4/3}}\biggl(R + UV^{2/3}\biggr) = 0 \eeq
In slightly more usual language the Hamiltonian constraint is
\beq \sigma^{ab}\sigma_{ab} - \frac{1}{6}\frac{(trp)^{2}}{V^{2}} - 
\frac{g}{V^{4/3}}\biggl(R - 16\pi\rho_{r}V^{2/3}\biggr) \eeq
where $\rho_{r}$ is the energy density of the radiation.
\subsection{Dust}
The Lagrangian in GR was
\beq \mathcal{L} = N\sqrt{g}\biggl(R + K^{ab}K_{ab} - (trK)^{2} + 
\rho_{d}\biggl(g^{\alpha\beta}U_{\alpha}U_{\beta} + 1\biggr)\biggr) \eeq
Conformalising this is a little less straightforward than the earlier cases. 
Firstly, what weight should we give to $\rho_{d}$? Well, Dicke prescribes that
under a conformal transformation of the form
\beq g_{ab} \gt \omega^{4}g_{ab} \eeq
we have
\beq \rho_{d} \gt \omega^{-2}\rho_{d} \eeq
Another point is that we wish the constraint
\beq g^{\alpha\beta}U_{\alpha}U_{\beta} + 1 = 0 \eeq
to hold in the new theory also and so we demand that
\beq \biggl(g^{\alpha\beta}U_{\alpha}U_{\beta} + 1\biggr) \gt 
\omega^{-4}\biggl(g^{\alpha\beta}U_{\alpha}U_{\beta} + 1\biggr) \eeq
Then the conformal Lagrangian is
\beq \mathcal{L} = N\psi^{2}\sqrt{g}\psi^{6}\biggl(\psi^{-4}\biggl(\rp\biggr) +
 \psi^{-4}S^{ab}S_{ab} - \psi^{-16}(trB)^{2} + 
\rho_{d}\psi^{-2}\biggl(g^{\alpha\beta}U_{\alpha}U_{\beta} + 
1\biggr)\psi^{-4}\biggr) \eeq
Simplifying we get
\beq \mathcal{L} = N\sqrt{g}\psi^{4}\biggl(\rp + S^{ab}S_{ab} - 
\psi^{-12}(trB)^{2} + 
\psi^{-2}\rho_{d}\biggl(g^{\alpha\beta}U_{\alpha}U_{\beta} + 1\biggr)\biggr) 
\eeq
The Hamiltonian constraint becomes
\beq \sigma^{ab}\sigma_{ab} - \frac{1}{6}\psi^{12}(trp)^{2} - 
g\psi^{8}\biggl(\rp - \psi^{-2}\rho_{d}\biggr) = 0 \eeq
where $\rho_{d}$ is the mass density of the dust.
In the compact without boundary case this becomes
\beq \sigma^{ab}\sigma_{ab} - \frac{1}{6}\frac{\psi^{12}(trp)^{2}}{V^{2}} - 
\frac{g\psi^{8}}{V^{4/3}}\biggl(\rp - \psi^{-2}\rho_{d}V^{1/3}\biggr) = 0 \eeq
In the physical representation this is
\beq \sigma^{ab}\sigma_{ab} - \frac{1}{6}\frac{(trp)^{2}}{V^{2}} - 
\frac{g}{V^{4/3}}\biggl(R - \rho_{d}V^{1/3}\biggr) = 0 \eeq
Phew! At last. These results will ALL be used when we consider the cosmological
 implications of the conformal theory and so that will all have been worth it 
soon.
\chapter{Four Dimensions!}
\section{Introduction}
Einstein's formulation of relativity, both special and general, was beautifully
 embodied in the $4$-dimensional spacetime. Although we have been dealing with 
a $(3+1)$-dimensional picture throughout it is possible for us to consider 
things in $4$-dimensions. The $4$-dimensional picture emerges very naturally 
in the static theory. As for the new theory, we do not attempt to find an 
action but deduce the field equations nonetheless.
\\ \\
Of course, at the end we will always have a breaking of full $4$-dimensional 
covariance due to the lapse fixing equations but it may be instructive to 
consider the $4$-dimensional picture even as a tool for making comparisons with
 GR.
\section{BOM Conformal Gravity}
In this section we construct a \emph{$4$}-dimensional action based on conformal
transformations of the \emph{$4$}-metric. We then decompose this to a 
$(3 + 1)$-dimensional form and from this we find the Jacobi action of the 
theory. Incredibly, it turns out to be the same as that of Barbour and \'{O} 
Murchadha.
\subsection{The Action}  
As given earlier (\ref{eh}) the Einstein-Hilbert action of general relativity 
is
\beq S = \int \sqrt{-^{(4)}g}\: \fourr \: d^{4}x \eeq 	
where $g_{\alpha\beta}$ is the 4-metric and $^{(4)}R$ is the four dimensional 
Ricci scalar. The action is varied with respect to $g_{\alpha\beta}$ and the 
resulting equations are the (vacuum) Einstein equations
\beq   G^{\alpha\beta} = \Biggl(R^{\alpha\beta} - 
\frac{1}{2}g^{\alpha\beta}R\Biggr) = 0   \eeq
We would like to construct an action which is invariant under \cts of the 
metric
\beq \fourg \longrightarrow \Omega^{2}\fourg \eeq
where $\Omega$ is a strictly positive function using the Einstein-Hilbert 
action as a guide. First we need to develop some machinery for dealing with 
conformal transformations.
\subsection{Dimensional Properties of Conformal Transformations}
A supposed problem with conformal transformations and different numbers of
dimensions is that various coefficients change when the number of dimensions
changes. This turns out not to be a problem in this analysis as will be shown.
\\ \\
Let us consider conformal transformations and the scalar curvature. If we make
a conformal transformation of the metric
\beq g_{\alpha\beta} \longrightarrow \Omega^{2}g_{\alpha\beta} \eeq
then the Ricci tensor transforms as
\beq
\begin{split} ^{(n)}R_{\alpha\beta} \longrightarrow &\:^{(n)}R_{\alpha\beta} 
+ 2(n - 2)\frac{(\grad_{\alpha}\Omega)\grad_{\beta}\Omega}{\Omega^{2}}
 - (n - 2) \frac{\grad_{\alpha}\grad_{\beta}\Omega}{\Omega} \\
&+ (3 - n)g_{\alpha\beta}\frac{(\grad_{\gamma}\Omega)\grad^{\gamma}\Omega}
{\Omega^{2}} - g_{\alpha\beta}\frac{\grad_{\gamma}\grad^{\gamma}\Omega}{\Omega}
 \end{split} \eeq
and the scalar curvature transforms as
\beq ^{(n)}R \longrightarrow \Omega^{-2}\biggl(\, ^{(n)}R - 
2(n-1)g^{\alpha\beta}\frac{\grad_{\alpha}\grad_{\beta}\Omega}{\Omega} + 
(n-1)(4-n)\frac{\grad_{\gamma}\Omega\grad^{\gamma}\Omega}{\Omega^{2}}\, 
\biggr) \eeq
where $n$ is the number of dimensions. A consequence is that the combination
\beq \phi^{2/s}\biggl(\, ^{(n)}R - 
\frac{4(n-1)}{(n-2)}g^{\alpha\beta}\frac{\grad_{\alpha} 
\grad_{\beta}\phi}{\phi}\,\biggr) \eeq 
is conformally invariant for any scalar function $\phi$ under the combined 
transformation
\beq g_{\alpha\beta} \longrightarrow \Omega^{2}g_{\alpha\beta} \: \: \: 
\text{,} \: \: \: \phi \longrightarrow \Omega^{s}\phi \eeq
where $s = 1 - \frac{n}{2}$. While this is true in any number of dimensions we
are of course most concerned with the $3$-dimensional and $4$-dimensional 
cases. In $3$ dimensions we have $s=-\frac{1}{2}$. Thus we get that
\beq \phi^{-4}\biggl(\; ^{(3)}R - \frac{8\grad^{2}\phi}{\phi}\;\biggr) \eeq
is conformally invariant under the transformation 
\beq g_{ab} \longrightarrow 
\Omega^{2}g_{ab} \:\:\: \text{,} \:\:\: \phi \longrightarrow 
\frac{\phi}{\sqrt{\Omega}} \eeq
In four dimensions $s = -1$ and the combination
\beq \phi^{-2}\biggl(\,\fourr - \frac{6\Box\phi}{\phi}\,\biggr) \eeq
is conformally invariant under the transformation \beq g_{\alpha\beta} 
\longrightarrow \Omega^{2}g_{\alpha\beta} \:\:\: \text{,} \:\:\: 
\phi\longrightarrow \frac {\phi}{\Omega} \eeq 
Then the combination
\beq \sqrt{-^{(4)}g}\phi^{2}\biggl(\,\fourr - \frac{6\Box\phi}{\phi}\,\biggr) 
\eeq is also conformally invariant.
This will be our Lagrangian density $\mathcal{L}$. Thus our action is
\beq  S = \int\mathcal{L}\: d^{4}x  \eeq
Before we decompose this to a $(3 + 1)$ form let us consider the 
$4$-dimensional structure and see what emerges.
\subsection{Varying with respect to $g_{\alpha\beta}$}
The variation with respect to $\fourg$ is quite straightforward. The resulting
equations are
\beq -\phi^{2}\biggl(R^{\alpha\beta} - \frac{1}{2}g^{\alpha\beta}R\biggr) 
+ 4 \grad^{\alpha}\phi\grad^{\beta}\phi - 
g^{\alpha\beta}\grad_{\gamma}\phi\grad^{\gamma}\phi - 2 \phi \grad^{\alpha} 
\grad^{\beta}\phi + 2 g^{\alpha\beta}\phi \Box \phi = 0  \eeq
This looks quite complicated but it is actually just
\beq \label{ceins} \overline{G^{\alpha\beta}} = 0 \eeq where 
$\overline{G^{\alpha\beta}}$ is just the Einstein tensor conformally 
transformed with conformal factor $\phi$. Equivalently, this is the Einstein 
tensor for the metric $\phi^{2}\fourg$. This interpretation will prove useful 
later.
\subsection{Varying with respect to $\phi$}
Again, this variation is fairly straightforward. We get
\beq \label{phi}\fourr  - \frac{6\Box\phi}{\phi} = 0\eeq
This is actually the trace of (\ref{ceins}) and so, as such, is redundant.
This can be viewed as a result of $\phi$ being pure gauge. The notion of free 
end-point variation of gauge variables \cite{jb} shows that for a pure gauge 
variable $\psi$ say, we may vary the action with respect to both $\psi$ and 
its time derivative $\dot{\psi}$ independently. Because $\phi$ is pure gauge 
here we may vary the action with respect to $\phi$ and $\dot{\phi}$ 
independently. This will be crucial in the theory. We shall return to this.
\subsection{A note on the action}
The form of the action as it stands is not conventional as it contains second 
time derivatives of the metric. However, the combination
\beq \fourr + 2A^{\alpha}_{\;\; ;\alpha}  \eeq
where $A^{\alpha} = \bigl(n^{\alpha}trK + a^{\alpha}\bigr)$ , $n^{\alpha}$ is 
the unit timelike normal and $a^{\alpha}$ is the four-acceleration of an 
observer travelling along $\mathbf{\underline{n}}$ contains no second time derivatives. 
(The coordinates $\alpha$ are general.) We write our Lagrangian as
\beq \mathcal{L} = \sqrt{-^{(4)}g}\phi^{2}
\biggl(\; \fourr + 2A^{\alpha}_{\;\; ;\alpha} - 2A^{\alpha}_{\;\; ;\alpha}
 - \frac{6\Box\phi}{\phi}\; \biggr) \eeq 
which then becomes
\beq \mathcal{L} = \sqrt{-^{(4)}g}
\biggl(\phi^{2}\biggl(\; \fourr + 2A^{\alpha}_{\;\; ;\alpha}\;\biggr) + 
4\phi\phi_{,\alpha}A^{\alpha} + 
6g^{\mu\nu}\grad_{\mu}\phi\grad_{\nu}\phi\biggr) \eeq
after some integration by parts. 
\\ \\
This Lagrangian contains no second time derivatives of the metric. Varying this
with respect to $\phi$ and $\dot{\phi}$ gives two conditions which combine to
give equation (\ref{phi}). Although we may do these variations here in a 
general coordinate form it will be more instructive to do a $(3+1)$-dimensional
decomposition and get the corresponding equations there.
\subsection{(3+1)-Decomposition}
We are now ready to consider the new action. This is
\beq S = \int \sqrt{-^{(4)}g}\phi^{2}\biggl(\,\fourr - 
\frac{6\Box\phi}{\phi}\,\biggr)d^{4}x \eeq
The $4$-dimensional scalar curvature decomposes as earlier. The action becomes
\beq S = \int \sqrt{-^{(4)}g}\phi^{2}\biggl(R - (trK)^{2} + \kk 
- 2A^{\alpha}_{\;\; ;\alpha} - 6\frac{\Box\phi}{\phi}\biggr) \; d^{4}x \eeq
Let's separate this into two terms $ S_{1} $ and $ S_{2} $ where,
\beq S_{1} = \int \sqrt{-^{(4)}g}\phi^{2}\biggl(R - (trK)^{2} + \kk 
- 2A^{\alpha}_{\;\; ;\alpha}\biggr) \; d^{4}x \eeq
and
\beq S_{2} = - \int 6\sqrt{-^{(4)}g}\phi\,\Box\phi\; d^{4}x \eeq
Consider the first term. In the ADM theory $A^{\alpha}_{\;\; ;\alpha}$ leads to
a total divergence which is discarded. However, the presence of the $\phi^{2}$
here changes this. Integrating by parts we get 
\beq -2\phi^{2}A^{\alpha}_{\;\; ;\alpha} \longrightarrow 
2(\phi^{2})_{;\alpha}A^{\alpha} \eeq
discarding the total divergence again. Decomposing this gives
\beq 
\begin{split} 2\Biggl(\dot{\phi^{2}}\biggl(n^{0}trK + a^{0}\biggr) + 
\biggl(\phi^{2}\biggr)_{,i}\biggl(n^{i}trK + a^{i}\biggr)\Biggr) = & 
4\phi\biggl(\dot{\phi}n^{0}trK + \phi_{,i}n^{i}trK\biggr) + 
4\phi\phi_{,i}a^{i} \\ = & \frac{4\phi}{N}\biggl(\dot{\phi} - 
\phi_{,i}N^{i}\biggr)trK + 
4\phi\phi_{,i}a^{i} \end{split} \eeq
using the fact that
\beq n^{\alpha} = \Bigl(1/N,-N^{m}/N\Bigr) \eeq
Then,
\beq
\begin{split} S_{1} = & \int N\sqrt{g}\phi^{2}\biggl(R - (trK)^{2} + 
\kk\biggr)\: dt\: d^{3}x 
\\ &  + \int 4\sqrt{g}\phi\Biggl[\biggl(\dot{\phi} - \phi_{,i}N^{i}\biggr)trK
 + N\phi_{,i}a^{i}\Biggr]\: dt\: d^{3}x \end{split} \eeq
We must now deal with $S_{2}$. This is,
\beq S_{2} = - \int 6\sqrt{-^{(4)}g}\phi\,\Box\phi\; d^{4}x \eeq
After a little integration by parts this is
\beq S_{2} = \int 6\sqrt{-^{(4)}g}\,g^{\mu\nu}\grad_{\mu}\phi\grad_{\nu}\phi 
d^{4}x \eeq
Decomposing this gives
\beq S_{2} =  \int 6N\sqrt{g} 
\Biggl(-\frac{1}{N^{2}}\biggl(\dot{\phi}\biggr)^{2} + 
\frac{2N^{i}}{N^{2}}\dot{\phi}\phi_{,i} + \biggl(g^{ij} - 
\frac{N^{i}N^{j}}{N^{2}}\biggr)\phi_{,i}\phi_{,j}\Biggr)\: dt\: d^{3}x \eeq
The full action is now
\beq
\begin{split} S = & \int N\sqrt{g}\phi^{2}\biggl(R - (trK)^{2} + \kk\biggr)\: 
dt\: d^{3}x \\ &+ \int 4\sqrt{g}\phi\Biggl[\biggl(\dot{\phi} - 
\phi_{,i}N^{i}\biggr)trK + N\phi_{,i}a^{i}\Biggr]\: dt\: d^{3}x \\ &  
+ \int 6N\sqrt{g}\Biggl(-\frac{1}{N^{2}}\biggl(\dot{\phi}\biggr)^{2} 
+ \frac{2N^{i}}{N^{2}}\dot{\phi}\phi_{,i} + \biggl(g^{ij} - 
\frac{N^{i}N^{j}}{N^{2}}\biggr)\phi_{,i}\phi_{,j}\Biggr)\: dt\: d^{3}x 
\end{split} \eeq
This looks like a much more complicated object than we began with. There will,
however, be much simplification. First, let's write it as
\beq
\begin{split} S = & \int N\sqrt{g}\phi^{2}\biggl(R - (trK)^{2} + \kk\biggr)\: 
dt\: d^{3}x \\ & + \int 4\sqrt{g}\phi\Biggl[\biggl(\dot{\phi} - 
\phi_{,i}N^{i}\biggr)trK + \grad_{i}\phi\grad^{i}N\Biggr]\: dt\: d^{3}x \\ & 
- \int \frac{6}{N}\sqrt{g}\biggl(\dot{\phi} - 
\phi_{,i}N^{i}\biggr)^{2}\: dt\: d^{3}x  + 
\int 6N\sqrt{g}\grad_{i}\phi\grad^{i}\phi \: dt\: d^{3}x  \end{split} \eeq
where we have used $ a^{i} = \frac{\grad^{i}N}{N} $. If we set 
$ \theta = -\frac{2}{\phi}\biggl(\dot{\phi} - \phi_{,i}N^{i}\biggr) $
then we get
\beq
\begin{split} S = & \int N\sqrt{g}\phi^{2}\biggl(R - (trK)^{2} + \kk\biggr)\: 
dt\: d^{3}x \\ & - \int 2\sqrt{g}\theta\phi^{2}trK\: dt\: d^{3}x - 
\int \frac{3}{2}\frac{\sqrt{g}\theta^{2}\phi^{2}}{N}\: dt\: d^{3}x \\ & + 
\int 6N\sqrt{g}\grad_{i}\phi\grad^{i}\phi \: dt\: d^{3}x + 
\int 4\sqrt{g}\phi\grad_{i}\phi\grad^{i}N\: dt\: d^{3}x \end{split} \eeq 
This becomes
\beq
\begin{split} S = & \int N\sqrt{g}\phi^{2}\biggl(R - (trK)^{2} + \kk\biggr)\: 
dt\: d^{3}x \\ &  - \int 2\sqrt{g}\theta\phi^{2}trK\: dt\: d^{3}x - 
\int \frac{3}{2}\frac{\sqrt{g}\theta^{2}\phi^{2}}{N}\: dt\: d^{3}x \\ & + \int 
2N\sqrt{g}\grad_{i}\phi\grad^{i}\phi \: dt\: d^{3}x - 
\int 4N\sqrt{g}\phi\grad^{2}\phi\: dt\: d^{3}x \end{split} \eeq
after some integration by parts. We notice that there might be a
possibility of ``completing some squares'' with terms involving K and those
involving $\theta$. We have,
\beq \label{ccfs} -N(trK)^{2} + N\kk -2\theta trK - 
\frac{3}{2}\frac{\theta^{2}}{N} \eeq
Let's try the combination,
\beq - N\biggl(trK + A\frac{\theta}{N}\biggr)^{2} + 
N\biggl(K_{ab} +B\frac{\theta g_{ab}}{N}\biggr)\biggl(K^{ab} + 
B\frac{\theta g^{ab}}{N}\biggr) \eeq
This gives us,
\beq - N(trK)^{2} - 2A\theta trK - A^{2}\frac{\theta^{2}}{N} + N\kk + 
2B\theta trK + 3B^{2}\frac{\theta^{2}}{N} \eeq
Comparing coefficients with equation (\ref{ccfs}) gives us,
\beq -2A + 2B = -2 \:\: \text{and} \: \: -A^{2} + 3B^{2} = -\frac{3}{2} \eeq
Solving here gives $ A = \frac{3}{2} $ and $ B = \frac{1}{2} $ and so we have,
\beq -N\biggl(trK + \frac{3}{2}\frac{\theta}{N}\biggr)^{2} + N\biggl(K_{ab} 
+ \frac{1}{2}\frac{\theta}{N}g_{ab}\biggr)\biggl(K^{ab} + 
\frac{1}{2}\frac{\theta}{N}g^{ab}\biggr)  \eeq
Finally, let us set  
\beq B_{ab} = \biggl(K_{ab} + \frac{\theta}{2N}g_{ab}\biggr) \eeq
Thus we get,
\beq - N\bigl(trB\bigr)^{2} + N\bb  \eeq
overall. Our full action is now,
\beq
\begin{split} S = & \int N\sqrt{g}\phi^{2}\biggl(R - (trB)^{2} + \bb\biggr)\: 
dt\: d^{3}x \\ & + \int 2N\sqrt{g}\grad_{i}\phi\grad^{i}\phi \: dt\: d^{3}x \\ 
& - \int 4N\sqrt{g}\phi\grad^{2}\phi \: dt\: d^{3}x \end{split} \eeq
We are now in a $(3 + 1)$-dimensional form and so we would like to use the 
power of $\phi$ which is appropriate in $3$ dimensions. From the earlier 
discussion of conformal invariance in different numbers of dimensions we find 
that we should use $\psi = \phi^{1/2}$ (or $\psi^{2} = \phi$). This is no more
than a relabelling to make things look neater. There is no real change to the 
theory in this mere relabelling. We get, 
\beq
\begin{split} S = & \int N\sqrt{g}\psi^{4}\biggl(R - (trB)^{2} + \bb\biggr)\: 
dt\: d^{3}x \\ & + \int 8N\sqrt{g}\psi^{2}\grad_{i}\psi\grad^{i}\psi \: dt\: 
d^{3}x - 
\int 8N\sqrt{g}\psi^{2}\grad_{i}\psi\grad^{i}\psi \: dt\: d^{3}x  \\ & - \int 
8N\sqrt{g}\psi^{3}\grad^{2}\psi \: dt\: d^{3}x \end{split} \eeq
Thus the action is
\beq \label{cgbk} S = \int N\sqrt{g}\psi^{4}\Biggl
(R - 8\frac{\grad^{2}\psi}{\psi} - (trB)^{2} + \bb\Biggr)\: dt\: d^{3}x \eeq
\\
This looks \emph{much} better! In fact, this is \emph{precisely} the action we 
found in Chapter 1 (\ref{admish})! By demanding $4$-dimensional conformal invariance we have
constructed the exact theory BOM found by demanding only $3$-dimensional 
conformal invariance.
\section{Conformally Related Solutions}
We should consider again the issue of conformally related solutions. Suppose we
 have a solution of the equations $g_{\alpha\beta}$ and $\phi$. If we perform a
 conformal transformation on this metric with conformal factor $\alpha$, say, 
the new metric $h_{\alpha\beta} = \alpha^{2}g_{\alpha\beta}$ must still be a 
solution. We find that the conformal factor this time is $\eta = 
\frac{\phi}{\alpha}$ and so we have $\phi^{2}g_{\alpha\beta} = 
\eta^{2}h_{\alpha\beta}$ which is yet another demonstration of the 
identification of conformally related solutions. In the physical representation
 the $4$-dimensional equations take the form of the Einstein equations in 
vacuum
\beq G^{\alpha\beta} = 0 \eeq
However, these are supplemented with the conformal conditions breaking the 
$4$-dimensional covariance and thus setting it apart from general relativity.
\subsection{Topological Considerations}
We found earlier that if the manifold was compact without boundary that we had 
to make a change to the action by adding in a volume term. Of course, with the 
introduction of the volume term we have a change in the original 
$4$-dimensional action also. This becomes,
\beq \int \frac{\sqrt{-^{(4)}g}\phi^{2}\biggl(\,\fourr - 
\frac{6\Box\phi}{\phi}\,\biggr)}{V(\phi)^{\frac{2}{3}}}\; d^{4}x \eeq
We have an implicit $(3+1)$ split here because $V$ is a purely 
three-dimensional quantity.We vary with respect to $^{(4)}g_{0\alpha}$ and 
$^{(4)}g_{ij}$ separately. (We vary with respect to the lower index case as 
$^{(4)}g_{ij} = g_{ij}$ and so both the numerator and the denominator may be 
varied with respect to the spatial part of the metric.) The variations give
\beq \overline{G^{0\alpha}} = 0 \eeq
and
\beq N\sqrt{g}\phi^{2}\overline{G^{ij}} + \frac{2}{3}g^{ij}\sqrt{g}C\phi^{3}
= 0 \eeq
where
\beq C = 
 \int \frac{N\sqrt{g}\psi^{4}\biggl(\,\rp\,\biggr)}{V(\psi)}d^{3}x \eeq
arises, as usual, due to variation of the volume. As earlier, 
$\overline{G^{\alpha\beta}}$ is the Einstein tensor of the metric 
$\phi^{2}g_{\alpha\beta}$ and $\psi^{2} = \phi$. We have used the Hamiltonian 
constraint to simplify $C$.
\\ \\
We can combine the equations to get
\beq \label{star} \overline{G^{\alpha\beta}} + 
\frac{2}{3N}h^{\alpha\beta}C\phi = 0 \eeq
where $h^{\alpha\beta}$ is the induced $3$-metric. This has the form
\begin{gather}
\begin{Vmatrix} h^{00} & h^{0k} \\ \\  h^{i0} & h^{ik} 
\end{Vmatrix} \quad
= \quad \begin{Vmatrix} \;\;0\;\; & \;\;0\;\; \\ \\  \;\;0\;\; & g^{ik}
\end{Vmatrix}
\end{gather}
We may lower the indices using $g_{\alpha\beta}$ to get
\begin{gather}
\begin{Vmatrix} h_{00} & h_{0k} \\ \\  h_{i0} & h_{ik} 
\end{Vmatrix} \quad
= \quad \begin{Vmatrix} N^{s}N_{s} & N_{k} \\ \\  N_{i} & g_{ik}
\end{Vmatrix}
\end{gather}
\\ \\
In the physical representation equation (\ref{star}) becomes
\beq \label{ces} G^{\alpha\beta} + \frac{2}{3N}h^{\alpha\beta}C = 0 \eeq
where now $C = \Big<NR\Big>$.
\\ \\
Of these ten equations, the four $0\alpha$ equations are identical to those in
general relativity while the remaining six differ by the new term which arose 
due to the variation of the volume. This new term is both time dependent and 
position dependent and so behaves like a ``non-constant cosmological 
constant.'' It will undoubtedly lead to new features, particularly in 
cosmology. However we shall not delve into this here. 
\\ \\
We must also do the variations with respect to $\phi$ and $\dot{\phi}$. The 
volume is independent of $\dot{\phi}$ and so this variation gives us exactly 
the same result as earlier, namely
\beq trB = 0 \eeq
However the volume is not independent of $\phi$ and so we will have a slight 
change. Varying with respect to $\phi$ gives us exactly what we found when we 
did the variation on the original form of the action (of course)
\beq N\psi^{3}\biggl(R - 7\frac{\grad^{2}\psi}{\psi}\biggr) - 
\grad^{2}\biggl(N\psi^{3}\biggr) = C\psi^{5}  \eeq
where 
\beq C = \int \frac{N\sqrt{g}\psi^{4}\biggl({\rp}\biggr)d^{3}x}{V(\psi)}\eeq
This becomes
\beq NR - \grad^{2}N = \Big<NR\Big> \eeq
in the physical representation.
\\ \\
Let us consider equation (\ref{ces}) again. Taking the trace gives us
\beq -^{(4)}R + 2\frac{\Big<NR\Big>}{N} = 0 \eeq
or
\beq N^{(4)}R = 2\Big<NR\Big> \eeq
If we average both sides of this equation we get
\beq \frac{\int N\sqrt{g}^{(4)}R d^{3}x}{\int \sqrt{g}d^{3}x} = 2\Big<NR\Big>
\eeq
Now, decomposing $^{(4)}R$ as earlier we get
\beq \int\frac{N\sqrt{g}\biggl(R - (trK)^{2} + \kk -2A^{\alpha}_{\;\; ;\alpha}
\biggr) d^{3}x}{V} = 2\Big<NR\Big> \eeq
This gives us
\beq \int\frac{N\sqrt{g}\biggl(R - (trK)^{2} + \kk\biggr)d^{3}x}{V} 
-2\int\frac{N\sqrt{g}A^{\alpha}_{\;\; ;\alpha} d^{3}x}{V} = 2\Big<NR\Big> 
\eeq
We notice that $N\sqrt{g} = \sqrt{-^{(4)}g}$ and so we may write
\beq \int\frac{N\sqrt{g}\biggl(R - (trK)^{2} + \kk\biggr)d^{3}x}{V} 
-2\int\frac{\sqrt{-^{(4)}g}A^{\alpha}_{\;\; ;\alpha} d^{3}x}{V} = 
2\Big<NR\Big> \eeq
Thus we have
\beq \int \frac{N\sqrt{g}\biggl(R - (trK)^{2} + \kk\biggr)d^{3}x}{V} 
-2\int\frac{\biggl(\sqrt{-^{(4)}g}A^{\alpha}\biggr)_{,\alpha} d^{3}x}{V} = 
2\Big<NR\Big> \eeq
The second term on the left hand side is a total $4$-divergence. We can discard
the spatial part to leave us with
\beq \int \frac{N\sqrt{g}\bigl(R - (trK)^{2} + \kk\bigr)d^{3}x}{V} 
-2\int\frac{\biggl(\sqrt{-^{(4)}g}A^{0}\biggr)_{,0} d^{3}x}{V} = 2\Big<NR\Big> 
\eeq
Using the Hamiltonian constraint we get
\beq \int\frac{\sqrt{g}\bigl(2NR\bigr) d^{3}x}{V} 
- 2\int\frac{\biggl(\sqrt{-^{(4)}g}A^{0}\biggr)_{,0} d^{3}x}{V} = 2\Big<NR\Big>
\eeq
This is 
\beq 2\Big<NR\Big> - 2\int\frac{\biggl(\sqrt{-^{(4)}g}A^{0}\biggr)_{,0} 
d^{3}x}{V} = 2\Big<NR\Big> \eeq
and so
\beq \int\biggl(\sqrt{-^{(4)}g}A^{0}\biggr)_{,0}d^{3}x = 0 \eeq
Thus
\beq \biggl(\sqrt{-^{(4)}g}A^{0}\biggr)_{,0} = 0 \eeq
But using the form of $A^{\alpha}$ which we gave earlier
\beq A^{\alpha} = \Bigl(n^{\alpha}trK + a^{\alpha}\Bigr) \eeq
we have
\beq \Biggl(\frac{\sqrt{-^{(4)}g}trK}{N}\Biggr)_{,0} = 0 \eeq
Recall once more that $\sqrt{^{(4)}g} = N\sqrt{g}$ to get
\beq \biggl(\sqrt{g}trK\biggr)_{,0} = 0 \eeq
which from the definition of $\pi^{ab}$ is
\beq \frac{\partial tr\pi}{\partial t} = 0 \eeq
Of course, this is already known from the propagation of the $tr\pi$ 
constraint. Thus we have demonstrated that there is no inconsistency in the 
equations. 
\\ \\
\underline{Note}: Although we have demonstrated this only in the physical 
representation it is equally valid in the general representation.
\\ \\
Of course, despite all these nice outcomes we know that the theory is flawed. 
Can we find a suitable $4$-dimensional picture for the new conformal theory?
\section{New Conformal Gravity}
Despite the natural way in which the $4$-dimensional picture emerged in the 
non-expanding theory we could have simply realised what the $4$-dimensional 
equations would have been by comparing with the GR cases. This is exactly how 
we will proceed here. (The change from the simple $\theta$ term in the original
 theory to the more complicated $\grad_{c}\xi^{c}$ makes finding an action much
 more difficult. Indeed whether anything is actually to be gained in finding a 
$4$-dimensional action remains to be seen.)
\subsection{Non-Compact}
Let's begin with the non-compact theory. We want to work with geometric 
quantities here rather than the Hamiltonian quantities. That is, in terms of the
 extrinsic curvature rather than the momentum. The three constraints are
\beq \label{one} A^{ab}A_{ab} - \frac{2}{3}(trK)^{2} - R = 0 \eeq
\beq \label{two} \grad_{b}(A^{ab} + \frac{2}{3}g^{ab}trK) = 0 \eeq
\beq \label{three} \grad_{c}trK = 0 \eeq
We also have the lapse-fixing equation
\beq \label{four} NR - \grad^{2}N +\frac{N(trp)^{2}}{4} = 0 \eeq
It is well known that the $0\alpha$ components of the Einstein tensor are the 
Hamiltonian and momentum constraints of GR. (Actually, this is true for the 
$G^{0}_{\alpha}$ components strictly speaking.) Let's consider the conformal 
constraints. The first two (\ref{one}) and (\ref{two}) are exactly the same as 
the GR constraints. We simply have
\beq G^{0\alpha} = 0 \eeq
as in GR. As for the $ij$ components, we have exactly the same as in GR also 
since the evolution equations are identical in both theories. Thus overall
\beq G^{\alpha\beta} = 0 \eeq
These are supplemented by the conformal constraints (\ref{three}) and 
(\ref{four}). These break the $4$-covariance. Nonetheless, the two theories are
 incredibly similar. Every solution of the conformal theory is a solution of GR
 although there are solutions of GR which are not solutions of the conformal 
theory (namely those which do not have a constant CMC slicing as prescribed by 
the conformal theory). When the volume terms are introduced it is not quite so 
simple.
\subsection{Compact Manifold}
As noted above, the constraints are the one up one down $0\alpha$ components of
 the Einstein tensor ($G^{0}_{\;\alpha}$). When both indices are raised we have
\beq \label{eone} G^{00} = - \frac{1}{2N^{2}}\biggl(A^{ab}A_{ab} - 
\frac{2}{3}(trK)^{2} - R\biggr) + 
\frac{N^{a}}{N^{2}}\grad_{b}\biggl(A_{a}^{\;b} - 
\frac{2}{3}g_{a}^{\;b}trK\biggr) \eeq
and
\beq \label{etwo} G^{0c} = \frac{N^{c}}{2N^{2}}\biggl(A^{ab}A_{ab} - 
\frac{2}{3}(trK)^{2} - R\biggr) + \grad_{b}\biggl(A^{cb} - 
\frac{2}{3}g^{cb}trK\biggr) - 
\frac{N^{c}N^{a}}{N^{2}}\grad_{b}\biggl(A_{a}^{\;b} - 
\frac{2}{3}g_{a}^{\;b}trK\biggr) \eeq
Although we have raised the indices both equations are still combinations of 
the constraints which are
\beq \label{one1} A^{ab}A_{ab} - \frac{2}{3}(trK)^{2}V^{2} - gR = 0 \eeq
\beq \label{two2} \grad_{b}\biggl(\frac{-A^{ab}}{V^{2/3}} + 
\frac{2}{3}g^{ab}trKV^{4/3}\biggr) = 0 \eeq
\beq \label{three3} \grad_{c}\biggl(trKV^{4/3}\biggr) = 0 \eeq
We also have the lapse-fixing equation
\beq \label{four4} NR - \grad^{2}N + N(trK)^{2}V^{2} = C \eeq
where $C = \Big<N\biggl(R + (trK)^{2}V^{2}\biggr)\Big>$.
\\ \\
The first two constraints (\ref{one1}) and (\ref{two2}) are very similar to the
 components of $G^{0\alpha}$. Indeed, just by adding and subtracting things we 
can get the constraints to appear explicitly as components of $G^{0\alpha}$ 
with extra terms. Consider the first constraint (\ref{one1}). This can be 
written as
\beq A^{ab}A_{ab} - \frac{2}{3}(trK)^{2} - gR + \frac{2}{3}(trK)^{2} 
- \frac{2}{3}(trK)^{2}V^{2} = 0 \eeq
That is
\beq A^{ab}A_{ab} - \frac{2}{3}(trK)^{2} - gR = \frac{2}{3}(trK)^{2}(V^{2} - 1)
\eeq
The second constraint (\ref{two2}) can be written as
\beq \label{two3} \frac{1}{V^{2/3}}\grad_{b}\biggl(-A^{ab} + 
\frac{2}{3}g^{ab}trKV^{2}\biggr) = 0 \eeq
Then
\beq \frac{1}{V^{2/3}}\grad_{b}\biggl(-A^{ab} + \frac{2}{3}g^{ab}trK\biggr) = 
\frac{1}{V^{2/3}}\frac{2}{3}\grad_{b}\biggl(g^{ab}trK(1 - V^{2})\biggr) \eeq
Thus, referring to the components of $G^{0\alpha}$ (\ref{eone}) and 
(\ref{etwo}) above we get
\beq G^{00} = \frac{1}{3N^{2}}(trK)^{2}(1 - V^{2}) + 
\frac{2}{3}\frac{N^{c}}{N^{2}}(V^{2} - 1)\grad_{c}trK \eeq
and
\beq G^{0c} = \frac{N^{c}}{3N^{2}}(trK)^{2}(V^{2} - 1) + 
\frac{2}{3}(V^{2} - 1)\biggl(g^{cd} - 
\frac{N^{c}N^{d}}{N^{2}}\biggr)\grad_{d}trK \eeq
Of course, by the first conformal constraint (\ref{three3}) we have 
$\grad_{c}trK = 0$. Thus we get
\beq G^{00} = \frac{1}{3N^{2}}(trK)^{2}(1 - V^{2}) \eeq
and
\beq G^{0c} = \frac{N^{c}}{3N^{2}}(trK)^{2}(V^{2} - 1) \eeq
We can further simplify (well, a little at least). Let's define
\beq 8\pi\rho_{ex} = \frac{1}{3N^{2}}(trK)^{2}(1 - V^{2}) \eeq
and
\beq 8\pi s_{\text{ex}}^{c} = \frac{N^{c}}{3N^{2}}(trK)^{2}(V^{2} - 1) \eeq
which gives
\beq s_{\text{ex}}^{c} = - N^{c}\rho_{\text{ex}} \eeq
We can now write
\beq G^{0\alpha} = 8\pi T_{\text{ex}}^{0\alpha} \eeq
where
\beq T_{\text{ex}}^{00} = \rho_{\text{ex}} \eeq
and
\beq T_{\text{ex}}^{0c} = s_{\text{ex}}^{c} \eeq
\\ \\
That was the easy part, so to speak. Finding the $ab$ components is more 
tricky. The most straightforward way to get this part is to look at the 
evolution equation for $\pi^{ab}$. For us, the most convenient way to express 
this is as
\beq \frac{\partial\pi^{ab}}{\partial t} = 
\frac{\partial\sigma^{ab}}{\partial t} + 
\frac{1}{3}\frac{\partial g^{ab}tr\pi}{\partial t} \eeq
This is
\beq \label{blah} \frac{\partial\pi^{ab}}{\partial t} = 
\frac{\partial\sigma^{ab}}{\partial t} + 
\frac{1}{3}tr\pi\frac{\partial g^{ab}}{\partial t} + 
\frac{1}{3}g^{ab}trp\frac{\partial \sqrt{g}}{\partial t} + 
\frac{1}{3}\sqrt{g}\frac{\partial trp}{\partial t} \eeq
Consider now each term on its own. There are quite a lot of calculations here 
which are all quite straightforward although cumbersome. We want to get each 
term expressed as the equivalent term in GR plus whatever extra terms there 
are. It is easiest explained by actually performing the calculations. The first
 term proceeds as follows. In Chapter $2$ we had
\beq
\begin{split} \frac{\partial\sigma^{ab}}{\partial t} = & - 
\frac{N\sqrt{g}}{V^{2/3}}\biggl(R^{ab} - \frac{1}{3}g^{ab}R\biggr) - 
\frac{2NV^{2/3}}{\sqrt{g}}\sigma^{ac}\sigma^{b}_{\;\;c} \\ & +
 \frac{\sqrt{g}}{V^{2/3}}\biggl(\grad^{a}\grad^{b}N - 
\frac{1}{3}g^{ab}\grad^{2}N\biggr) + \grad_{c}\biggl(\sigma^{ab}N^{c}\biggr) \\
 & - \sigma^{bc}\grad_{c}N^{a} - 
\sigma^{ac}\grad_{c}N^{b} + \frac{N}{3\sqrt{g}V^{4/3}}\sigma^{ab}tr\pi 
\end{split} \eeq
Let's factor out $\frac{1}{V^{2/3}}$. We get
\beq
\begin{split} \frac{\partial\sigma^{ab}}{\partial t} = & 
\frac{1}{V^{2/3}}\biggl(- 
N\sqrt{g}\biggl(R^{ab} - \frac{1}{3}g^{ab}R\biggr) - 
\frac{2NV^{4/3}}{\sqrt{g}}\sigma^{ac}\sigma^{b}_{\;\;c} \\ & +
 \sqrt{g}\biggl(\grad^{a}\grad^{b}N - \frac{1}{3}g^{ab}\grad^{2}N\biggr) + 
V^{2/3}\grad_{c}\biggl(\sigma^{ab}N^{c}\biggr) \\ & - 
V^{2/3}\sigma^{bc}\grad_{c}N^{a} - V^{2/3}\sigma^{ac}\grad_{c}N^{b} \\ & + 
\frac{N}{3\sqrt{g}V^{2/3}}\sigma^{ab}tr\pi\biggr) \end{split} \eeq
Now we need to change momenta to curvatures. Recall that we had (\ref{oops})
\beq \sigma^{ab} = - \sqrt{g}\frac{S^{ab}}{V^{2/3}} \eeq
and
\beq tr\pi = 2\sqrt{g}(trK)V^{4/3} \eeq
We get
\beq
\begin{split} \frac{\partial\sigma^{ab}}{\partial t} = & 
\frac{1}{V^{2/3}}\biggl(- N\sqrt{g}\biggl(R^{ab} - \frac{1}{3}g^{ab}R\biggr) - 
\frac{2NgV^{4/3}}{\sqrt{g}}\frac{S^{ac}S^{b}_{\;\;c}}{V^{4/3}} \\ & +
 \sqrt{g}\biggl(\grad^{a}\grad^{b}N - \frac{1}{3}g^{ab}\grad^{2}N\biggr) - 
V^{2/3}\grad_{c}\biggl(\frac{\sqrt{g}S^{ab}}{V^{2/3}}N^{c}\biggr) \\ & +
 V^{2/3}\frac{\sqrt{g}S^{bc}}{V^{2/3}}\grad_{c}N^{a} 
+ V^{2/3}\frac{\sqrt{g}S^{ac}}{V^{2/3}}\grad_{c}N^{b} \\ & - 
\frac{2Ng}{3\sqrt{g}V^{2/3}}\frac{S^{ab}}V^{2/3}trKV^{4/3}\biggr) \end{split} 
\eeq
Simplifying we get
\beq
\begin{split} \frac{\partial\sigma^{ab}}{\partial t} = & 
\frac{1}{V^{2/3}}\biggl(- N\sqrt{g}\biggl(R^{ab} - \frac{1}{3}g^{ab}R\biggr) - 
2N\sqrt{g}S^{ac}S^{b}_{\;\;c} \\ & +
 \sqrt{g}\biggl(\grad^{a}\grad^{b}N - \frac{1}{3}g^{ab}\grad^{2}N\biggr) - 
\sqrt{g}\grad_{c}\biggl(S^{ab}N^{c}\biggr) \\ & +
\sqrt{g}S^{bc}\grad_{c}N^{a} + \sqrt{g}S^{ac}\grad_{c}N^{b} \\ & -
\frac{2N\sqrt{g}}{3}S^{ab}trK\biggr) \end{split} \eeq
The quantity inside the outer brackets is exactly what we have in GR.
\beq \biggl(\frac{\partial\sigma^{ab}}{\partial t}\biggr)_{\text{CG}} = 
\frac{1}{V^{2/3}}\biggl(\frac{\partial\sigma^{ab}}{\partial t}\biggr)_{\text{GR}}\eeq
This shows the correspondence between the two theories as closely as possible 
and there are no extra terms. Considering (\ref{blah}) again and taking the 
second term on the right hand side we get
\beq 
\begin{split} \frac{1}{3}tr\pi\frac{\partial g^{ab}}{\partial t} = & - 
\frac{1}{3}tr\pi g^{ac}g^{bd}\frac{\partial g_{cd}}{\partial t} \\ = & 
- \frac{1}{3}tr\pi\biggl(\frac{2NV^{2/3}}{\sqrt{g}}\sigma^{ab} - 
\frac{1}{3}\frac{Ntr\pi g^{ab}V^{2/3}}{V^{2}} + (KN)^{ab}\biggr) \\ = & - 
\frac{2NV^{2/3}}{3\sqrt{g}}\sigma^{ab}tr\pi + 
\frac{N(tr\pi)^{2}}{9\sqrt{g}V^{4/3}} - \frac{}{3}tr\pi(KN)^{ab} \\ = & 
\frac{4N}{3}\sqrt{g}S^{ab}trKV^{4/3} + \frac{4N}{9}\sqrt{g}(trK)^{2}V^{4/3} - 
\frac{2}{3}\sqrt{g}trKV^{4/3}(KN)^{ab} \\ = & 
\frac{1}{V^{2/3}}\biggl(\frac{4N}{3}\sqrt{g}S^{ab}trKV^{2} + 
\frac{4N}{9}\sqrt{g}(trK)^{2}V^{2}  - 
\frac{2}{3}\sqrt{g}trKV^{2}(KN)^{ab}\biggr) \\ = & 
\frac{1}{V^{2/3}}\biggl(- 
\frac{1}{3}\biggl(tr\pi g^{ac}g^{bd}\frac{\partial g_{cd}}{\partial t}\biggr)
_{\text{GR}} + \frac{4N}{3}\sqrt{g}S^{ab}trK(V^{2} - 1) \\ & + 
\frac{4N}{9}\sqrt{g}(trK)^{2}(V^{2} - 1) - 
\frac{2}{3}\sqrt{g}trK(V^{2} - 1)(KN)^{ab}\biggr)
\end{split} \eeq
This shows the relationship between the two theories as closely as possible for
 this term. The third term in (\ref{blah}) is
\beq 
\begin{split} \frac{1}{6}g^{ab}tr\pi g^{cd}\frac{\partial g_{cd}}{\partial t} =
& \frac{1}{6}g^{ab}tr\pi\biggl(- \frac{Ntr\pi}{\sqrt{g}V^{4/3}} + 
2\grad_{c}N^{c}\biggr) \\ = & - \frac{N}{6\sqrt{g}V^{4/3}}g^{ab}(tr\pi)^{2} + 
\frac{1}{3}g^{ab}tr\pi\grad_{c}N^{c} \\ = & 
\frac{1}{V^{2/3}}\biggl(- \frac{2}{3}N\sqrt{g}g^{ab}(trK)^{2}V^{2} + 
\frac{2}{3}\sqrt{g}g^{ab}trKV^{2}\grad_{c}N^{c}\biggr) \\ = & 
\frac{1}{V^{2/3}}\biggl(
\biggl(\frac{1}{6}g^{ab}tr\pi g^{cd}\frac{\partial g_{cd}}{\partial t}\biggr)
_{\text{GR}} - \frac{2}{3}N\sqrt{g}g^{ab}(trK)^{2}(V^{2} - 1) \\ & + 
\frac{2}{3}\sqrt{g}g^{ab}trK(V^{2} - 1)\grad_{c}N^{c}\biggr) \end{split} \eeq
The final term in (\ref{blah}) is the most straightforward. We have
\beq \frac{1}{3}g^{ab}\sqrt{g}\frac{\partial trp}{\partial t} = 
\frac{1}{V^{2/3}}\biggl(\frac{1}{3}\biggl(g^{ab}\sqrt{g}
\frac{\partial trp}{\partial t}\biggr)_{\text{GR}} + 
\frac{2}{3}N\sqrt{g}(trK)^{2}(V^{2} - 1) - \frac{2}{3}\sqrt{g}g^{ab}C\biggr) 
\eeq
where $C = \Big<N\biggl(R + (trK)^{2}V^{2}\biggr)\Big>$.
The overall result is
\beq
\begin{split} \biggl(\frac{\partial\pi^{ab}}{\partial t}\biggr)_{\text{CG}} = 
& \frac{1}{V^{2/3}}
\biggl[\biggl(\frac{\partial\pi^{ab}}{\partial t}\biggr)_{\text{GR}} + 
\frac{4}{3}N\sqrt{g}S^{ab}trK(V^{2} - 1) + 
\frac{4}{9}N\sqrt{g}g^{ab}(trK)^{2}(V^{2} - 1) \\ & + 
\frac{2}{3N}\sqrt{g}trK(V^{2} - 1)\biggl(g^{ab}\grad_{c}N^{c} - 
(KN)^{ab}\biggr) - \frac{2}{3}\sqrt{g}g^{ab}C\biggr] \end{split} \eeq
Of course,
\beq \biggl(\frac{\partial\pi^{ab}}{\partial t}\biggr)_{\text{GR}} = 
\frac{\delta\mathcal{L}_{\text{GR}}}{\delta g_{ab}} = - N\sqrt{g}G^{ab} \eeq
Thus we get
\beq
\begin{split} G^{ab} = & \frac{4}{3}S^{ab}trK(V^{2} - 1) + 
\frac{4}{9}g^{ab}(trK)^{2}(V^{2} - 1) \\ & + 
\frac{2}{3N}trK(V^{2} - 1)\biggl(g^{ab}\grad_{c}N^{c} - (KN)^{ab}\biggr) - 
\frac{2}{3N}g^{ab}C \end{split} \eeq
Let's label the right hand side as $8\pi s^{ab}_{\text{ex}}$. Thus we have
\beq G^{\alpha\beta} = 8\pi T^{\alpha\beta}_{\text{ex}} \eeq
where $T^{00}_{\text{ex}} = \rho_{\text{ex}}$, $T^{0i}_{\text{ex}} = 
s^{i}_{\text{ex}}$ and $T^{ab}_{\text{ex}} = s^{ab}_{\text{ex}}$. Here we are 
assigning an energy-momentum tensor to the expansion.
\\ \\
We could change notation here and instead of using an energy-momentum tensor 
relabel again. We'll define 
\beq - C^{\alpha\beta} = 8\pi T^{\alpha\beta}_{\text{ex}} \eeq
The field equations are
\beq G^{\alpha\beta} - C^{\alpha\beta} = 0 \eeq
In both cases, of course, the equations are supplemented by the conformal 
constraint
\beq \grad_{c}trK = 0 \eeq
and the lapse-fixing equation
\beq NR - \grad^{2}N + N(trK)^{2}V^{2} = C \eeq
\section{Special Case}
A special case of the theory is when $trK = 0$. We would expect the theory to 
reduce to the original static theory. If we look at the $4$-dimensional 
equations we see that $trK = 0$ gives exactly (\ref{ces})
\beq G^{\alpha\beta} + \frac{2}{3N}h^{\alpha\beta}C = 0 \eeq
where now $C = \Big<NR\Big>$. The conformal constraint and lapse-fixing 
equations also reduce to those of the static theory. We could check this in any
 of the various formulations we have considered and it would work out in each 
and every one.
\section{The Solar System}
A necessary result for any theory of gravity is that it reproduce the 
well-tested (talk about an understatement!) solar system results. The field 
equations as presented here offer a good chance to do just this. If we take the
 solar system to be isolated and asymptotically flat then clearly we must use 
the asymptotically flat version of this theory. In that case we have complete 
agreement. The maximally sliced solar system result of GR is well known and 
thus we reproduce the results.
\\ \\
However, we could treat the solar system as part of a larger solution. Clearly 
then we cannot treat it as an isolated, asymptotically flat system. However, we
 still assume it to be static. Then we have $trK = 0$. We have just seen that 
the only difference between our field equations and those of GR for a static 
region is the term $\frac{2}{3N}h^{\alpha\beta}C$ where $C = \Big<NR\Big>$. The
$0\alpha$ equations are exactly the same here. Indeed, if $C$ is small. In a 
closed FRW universe this term will certainly be very small as the curvature is 
very small. Thus we recover the solar system results quite easily.
\section{Comment}
It is very interesting that the $4$-dimensional picture is so similar to the 
standard GR picture. Of course, the $(3+1)$ form is also very similar to that
 of GR but when we move to strictly geometrical quantities the similarities 
show up all the more so. We could do this in the $(3+1)$ formalism also by 
replacing the momentum terms with their corresponding curvature terms. However,
 doing this a second time (since we have done it here already) seems excessive.
\chapter{Cosmology}
\section{Introduction}
Despite the successes enjoyed by the original theory, it suffers from the fact 
that it predicts a static universe and so expansion is automatically 
prohibited. As a result, all of the successes of the big bang picture are lost 
and in particular the cosmological redshift - an experimental fact - is 
unexplained. In the new theory we have succeeded in recovering expansion and 
with this renewed confidence we should examine further the cosmological 
implications of the new theory. To begin with it will be instructive to briefly 
review cosmology $\grave{\text{a}}$ la GR.
\section{Cosmology In General Relativity}
In GR the dynamics are all in the Hamiltonian constraint. This is written in 
terms of the geometry and sources in question and from this the cosmological 
dynamics of the universe are determined. We'll assume the standard 
Friedmann-Robertson-Walker (FRW) metrics and consider each in turn beginning 
with the open universe.
\subsection{Open Universe}
The FRW metric for the open universe is
\beq d\sigma^{2} = a(t)^{2}\biggl[d\chi^{2} + \text{sinh}^{2}\chi(d\theta^{2} +
\text{ sin}^{2}\theta d\phi^{2})\biggr] \eeq
The tracefree part of the extrinsic curvature, $A_{ab}$, is zero and the trace 
part is $-\frac{3\dot{a}}{a}$. Then
\beq trp = 2trK = -\frac{6\dot{a}}{a} \eeq
The Hamiltonian constraint is
\beq R + \frac{1}{6}(trp)^{2} = 0 \eeq
In cosmological terms this becomes
\beq - \frac{6}{a^{2}} +\frac{6\dot{a}^{2}}{a^{2}} = 0 \eeq
This can be solved easily to give
\beq a = t + a_{i} \eeq
where $a_{i}$ is the value of $a$ at $t = 0$. Let's move to the flat case.
\subsection{Flat Universe}
The FRW metric for a flat universe is given by
\beq d\sigma^{2} = a(t)^{2}\biggl[d\chi^{2} + \chi^{2}(d\theta^{2} + 
\text{sin}^{2}\theta d\phi^{2})\biggr] \eeq
The extrinsic curvature is unchanged from the above case. The Hamiltonian 
constraint is now
\beq 0 + \frac{6\dot{a}^{2}}{a^{2}} = 0 \eeq
and so we get
\beq \dot{a} = 0 \eeq
a static universe.
\\ \\
Of course, we haven't tried adding matter to the system in either of these two 
cases. We could do this easily and get different results for $a(t)$. The whole
point here is to compare (and contrast) the predictions of the two theories and
 for reasons that will become apparent soon, we need only concern ourselves 
with the closed universe in any detail.
\subsection{Closed Universe}
The FRW metric for the closed universe is
\beq d\sigma^{2} = a(t)^{2}\biggl[d\chi^{2} + 
\text{sin}^{2}\chi^{2}(d\theta^{2} + \text{sin}^{2}\theta d\phi^{2})\biggr] 
\eeq
The extrinsic curvature is the same yet again.This time the Hamiltonian 
constraint is
\beq R + \frac{1}{6}(trp)^{2} = 0 \eeq
\beq \frac{6}{a^{2}} + \frac{6\dot{a}^{2}}{a^{2}} = 0 \eeq
Thus
\beq \dot{a}^{2} = - 1 \eeq
which is a contradiction. Thus we \emph{cannot} have a closed and matter-free 
FRW universe. Let's try adding matter. We will try three different types, 
namely a cosmological constant, radiation and dust.
\subsubsection{Cosmological Constant}
With the introduction of a cosmological constant the Hamiltonian constraint 
becomes
\beq R + \frac{1}{6}(trp)^{2} = \Lambda \eeq
and so
\beq \frac{6}{a^{2}} + \frac{6\dot{a}^{2}}{a^{2}} = \Lambda \eeq
and then
\beq \dot{a}^{2} = \frac{\Lambda a^{2}}{6} \eeq
We can now solve this for $a(t)$ (if we wish).
\subsubsection{Radiation}
Radiation couples to gravity in GR in the Hamiltonian constraint as
\beq R - \sigma^{ab}\sigma_{ab} + \frac{1}{6}(trp)^{2} = 16\pi\rho_{r} \eeq
where $\rho_{r}$ is the radiation energy density. For radiation
\beq \rho_{r} = \rho_{r_{0}}\frac{a_{0}^{4}}{a^{4}} \eeq
where a subscript $0$ denotes the value today. Thus the Hamiltonian constraint 
becomes
\beq \frac{6}{a^{2}} + \frac{6\dot{a}^{2}}{a^{2}} = 
16\pi\rho_{r_{0}}\frac{a_{0}^{4}}{a^{4}} \eeq
Again, we can solve this for $a(t)$.
\subsubsection{Dust}
Dust couples to gravity in the Hamiltonian constraint as
\beq R - \sigma^{ab}\sigma_{ab} + \frac{1}{6}(trp)^{2} = 16\pi\rho_{d} \eeq
where $\rho_{d}$ is the dust energy density. For dust
\beq \rho_{d} = \rho_{d_{0}}\frac{a_{0}^{3}}{a^{3}} \eeq
where a subscript $0$ again denotes the value today. Thus the Hamiltonian 
constraint becomes
\beq \frac{6}{a^{2}} + \frac{6\dot{a}^{2}}{a^{2}} = 
16\pi\rho_{d_{0}}\frac{a_{0}^{3}}{a^{3}} \eeq
Again, we can solve this for $a(t)$.
\\ \\
Although we have glossed over some details and not solved explicitly for $a(t)$ 
each time, the important features should be clear. Let's move to the conformal 
theory and see what happens there.
\section{Cosmology in the Conformal Theory}
In the conformal theory we will need to consider \emph{two} constraints. The 
Hamiltonian constraint and the conformal constraint $trp \equiv 
\text{constant}$ will both have significance. Again, we will assume the 
standard FRW metrics and consider each in turn. We will use the results for the
 Hamiltonian constraints from the last chapter in every case.
\subsection{Open Universe}
The metric in this case is
\beq d\sigma^{2} = a(t)^{2}\biggl[d\chi^{2} + \text{sinh}^{2}\chi(d\theta^{2} +
\text{ sin}^{2}\theta d\phi^{2})\biggr] \eeq
\\
The first thing we should consider is the constant $trp$ constraint. The 
extrinsic curvature is of course the same as in GR: the tracefree part, 
$A_{ab}$, is zero and we get $trK = -\frac{3\dot{a}}{a}$. Thus, the equation to
 be solved is
\beq \int \frac{1}{a}\;\;da = \int C\;\;dt \eeq
This gives us
\beq \label{prob} a = Ae^{Bt} \eeq
Let's examine the Hamiltonian constraint now.
\\ \\
The Hamiltonian constraint here is
\beq R + \frac{(trp)^{2}}{6} = 0 \eeq
Thus we get that
\beq R = -\frac{(trp)^{2}}{6} \eeq
that is, $R$ is constant. Thus, $a$ must be constant. This is not a great 
result as we have again lost expansion. Let's try to couple matter to the 
system.
\subsubsection{Cosmological Constant}
Adding in a cosmological constant here does not change the essence of the above
 result. We still have no expansion. The next obvious matter to try is dust.
\subsubsection{Dust}
Dust couples just as in GR to give a Hamiltonian constraint
\beq R + \frac{(trp)^{2}}{6} = \rho \eeq
In cosmological terms we get
\beq -\frac{6}{a^{2}} + \frac{(trp)^{2}}{6} = \frac{\rho_{0}a_{0}^{3}}{a^{3}} 
\eeq
Clearly, the solution found above (\ref{prob}) does not work here unless, 
again, $a$ is constant in time. Thus, we have lost expansion again. We could 
attempt to couple in radiation but we would find the same problem arising. The 
open FRW universe seems a lost cause. Let's move on.
\subsection{Flat Universe}
The metric in this case is
\beq d\sigma^{2} = a(t)^{2}\biggl[d\chi^{2} + \chi^{2}(d\theta^{2} + 
\text{sin}^{2}\theta d\phi^{2})\biggr] \eeq
Applying the constant $trp$ constraint gives us the same as in the open 
universe case.
\beq a = Ae^{Bt} \eeq
Without wasting any more time, it isn't too difficult to see that we will come
up against the very same problems as in the open universe. Thus, both the 
standard open and flat FRW universes seem to be lost causes. This is worrying. 
Will this trend continue? Will the theory fail yet again? Let's find out.
\subsection{The Closed Universe}
The metric here is
\beq d\sigma^{2} = a(t)^{2}\biggl[d\chi^{2} + \text{sin}^{2}\chi(d\theta^{2} + 
\text{sin}^{2}\theta d\phi^{2})\biggr] \eeq
We find $K_{ab}$ and $trK$ as usual. Yet again we get
\beq A_{ab} = 0 \eeq
and
\beq trK = -\frac{3\dot{a}}{a} \eeq
However, in this case we have a totally new expression (\ref{oops}) for finding
 the momentum
\beq trp = 2(trK)V^{4/3} \eeq
The volume of a hypersurface in this universe is
\beq V = 2\pi^{2}a^{3} \eeq
Thus we get
\beq
\begin{split} trp = &  
2\biggl(-\frac{3\dot{a}}{a}\biggr)(2\pi^{2})^{4/3}a^{4} 
\\ = & -6(2\pi^{2})^{4/3}a^{3}\dot{a} \end{split} \eeq
The constraint becomes
\beq -6(2\pi)^{4/3}a^{3}\dot{a} = D \eeq
where $D$ is a constant. Integrating across we get
\beq -6(2\pi)^{4/3}\int a^{3}\;\;da = \int D\;\;dt \eeq
Thus we get an expression for $a(t)$.
\beq \label{at1} a^{4} = Ct + a_{i}^{4} \eeq
where $C$ is a constant both spatially and temporally and $a_{i}$ is the 
radius of the universe at $t = 0$.
\\ \\
This must hold regardless of what matter we include which is precisely the 
point which caused the other cases to fail. What will happen here?
\subsubsection{Matter free}
Let's first see what happens if there is no matter. The Hamiltonian constraint
 is
\beq R + \frac{1}{6}\frac{(trp)^{2}}{V^{2/3}} = 0 \eeq
We know that $R = \frac{6}{a^{2}}$ and so we get
\beq \frac{6}{a^{2}} + \frac{1}{6}\frac{(trp)^{2}}{(2\pi^{2})^{2/3}a^{2}}= 0 
\eeq
The $a^{2}$ cancels in each term and we get
\beq 6 + \frac{1}{6}\frac{(trp)^{2}}{(2\pi^{2})^{2/3}} = 0 \eeq
We get an identity for $trp$ \emph{consistent} with the constant $trp$ 
constraint! Of course, this particular identity cannot hold for real $trp$ but 
it is a step in the right direction at least. Let's try some matter. We will 
consider three types of matter here: a cosmological constant, dust and
 radiation. These have been treated with regard to the conformal theory 
in an earlier chapter but the beauty of what happens (from a cosmological point
 of view) was not noticed. The various Hamiltonian constraints which we have 
found earlier will be examined. Of course, we will work purely in the physical 
representation.
\subsubsection{Cosmological Constant}
The Hamiltonian constraint here is
\beq R + \frac{1}{6}\frac{(trp)^{2}}{V^{2/3}} + \frac{\Lambda}{V^{2/3}} = 0 
\eeq
Substituting in the quantities in terms of $a$ we get
\beq \frac{6}{a^{2}} + \frac{1}{6}\frac{(trp)^{2}}{(2\pi^{2})^{2/3}a^{2}} + 
\frac{\Lambda}{(2\pi^{2})^{2/3}a^{2}} = 0 \eeq
and yet again the $a^{2}$ cancels across the entire expression to give
\beq (trp)^{2} = -6\Lambda - 36(2\pi^{2})^{2/3} \eeq
This is entirely consistent with the earlier expression for $a(t)$. Also, we 
notice that since both sides must be positive we get a condition on $\Lambda$
\beq \Lambda \leqslant -6(2\pi^{2})^{2/3} \eeq
This is very encouraging. Let's try electromagnetism.
\subsubsection{Electromagnetism}
The Hamiltonian constraint here was found to be
\beq \sigma^{ab}\sigma_{ab} - \frac{1}{6}\frac{(trp)^{2}}{V^{2}} + 
\frac{\mu^{i}\mu_{i}}{V^{2/3}} - \frac{U}{V^{2/3}} -\frac{gR}{V^{4/3}} = 0 \eeq
Moving things about a little gives us
\beq R - \frac{\sigma_{ab}\sigma^{ab}V^{4/3}}{g} + 
\frac{1}{6}\frac{(trp)^{2}}{V^{2/3}} = 
\biggl(\mu_{i}\mu^{i} - U\biggr)\frac{V^{2/3}}{g} \eeq
This becomes
\beq R - \frac{\sigma_{ab}\sigma^{ab}V^{4/3}}{g} + 
\frac{1}{6}\frac{(trp)^{2}}{V^{2/3}} = 16\pi\rho_{r}V^{2/3} \eeq
where $\rho_{r}$ is the energy density of radiation in the universe. For 
radiation we have that
\beq \rho_{r} = \frac{\rho_{r_{0}}a_{0}^{4}}{a^{4}} \eeq
where $\rho_{r_{0}}$ is the energy density of radiation today and $a_{0}$ is 
the radius of the universe today. The constraint becomes
\beq \frac{6}{a^{2}} + \frac{(trp)^{2}}{6(2\pi)^{2/3}a^{2}} = 
16\pi\frac{\rho_{r_{0}}a_{0}^{4}}{a^{4}}(2\pi)^{2/3}a^{2} \eeq
Yet again, the $a^{2}$ cancels right across the 
board and we get an identity involving the collective ``energies'' which is
 completely devoid of dynamical content, namely
\beq 6 + \frac{(trp)^{2}}{6(2\pi)^{2/3}} = 
28^{2/3}\pi^{5/3}\rho_{r_{0}}a_{0}^{4} \eeq
This is completely consistent with the solution of $a(t)$ from earlier. In 
fact, just like with the cosmological constant we can get the total ``radiation
 energy'' in the universe in terms of $trp$ (or vice versa). Let's now consider
 dust.
\subsubsection{Dust}
Our Hamiltonian constraint here is
\beq R - \sigma^{ab}\sigma_{ab}V^{4/3} + \frac{1}{6}(trp)^{2}V^{-2/3} - 
16\pi\rho_{d}V^{1/3} = 0 \eeq
where $\rho_{d}$ is the energy density of dust. For dust we have that
\beq \rho_{d} = \frac{\rho_{d_{0}}a_{0}^{3}}{a^{3}} \eeq
where $\rho_{d_{0}}$ is the energy density of dust today and $a_{0}$ is the 
radius of the universe today. In cosmological terms this becomes
\beq \frac{6}{a^{2}} + \frac{1}{6}\frac{(trp)^{2}}{(2\pi)^{2/3}a^{2}} = 
16\pi\rho_{d_{0}}\frac{a_{o}^{3}}{a^{3}}(2\pi)^{1/3}a \eeq
Yet again, all the $a$ terms cancel and we get
\beq 6 + \frac{1}{6}\frac{(trp)^{2}}{(2\pi)^{2/3}} = 
8192^{1/3}\pi^{4/3}\rho_{d_{0}}a_{0}^{3} \eeq
Thus, yet again, it is an identity involving strictly non-time evolving terms.
 The dynamical content is still in $trp = \text{constant}$.
\subsubsection{General Case}
We can put all the results together here to get
\beq 6 + \frac{1}{6}\frac{(trp)^{2}}{(2\pi)^{2/3}} = 
8192^{1/3}\pi^{4/3}\rho_{m_{0}}a_{0}^{3} + 
128^{2/3}\pi^{5/3}\rho_{r_{0}}a_{0}^{4} + \frac{\Lambda}{(2\pi)^{2/3}} \eeq
(Here, $\rho_{m}$ is the matter mass density in the universe. It behaves just 
like dust from the point of view of the Hamiltonian constraint.) However, the 
evolution of the universe is still governed by the equation found earlier
\beq a^{4} = Dt + a_{i}^{4} \eeq
We get an ever-expanding decelerating universe. The Hamiltonian constraint seems
 to have been promoted to an identity for the various energies. This needs 
further examination.
\\ \\
(Note: We should note that although we have chosen to treat the constant $trp$ 
condition separately from the Hamiltonian constraint that in the closed 
universe they amount to the same thing and the form of the Hamiltonian 
constraint actually \emph{determines} that constant!)
\section{Cosmological Parameters}
Very often cosmological scenarios are described using various parameters. The 
most important two being the Hubble parameter and the deceleration parameter. 
These are defined as
\beq H = \frac{1}{3}\frac{d}{dt}(\text{ln}\sqrt{g}) \eeq
for the Hubble parameter and
\beq q = - \frac{\ddot{a}}{a}\frac{1}{H^{2}} \eeq
for the deceleration parameter. For the FRW metrics the Hubble parameter is 
just
\beq H = \frac{\dot{a}}{a} \eeq
What does the conformal theory say about these?
\subsection{Hubble Parameter}
We'll take the same definition for the Hubble parameter
\beq H = \frac{1}{3}\frac{d ln\sqrt{g}}{dt} \eeq
At first glance, with all the volume terms which have been introduced to the 
theory it seems unlikely that the result will be the same. It is not too 
difficult to go through the calculation to find that, indeed, the result does 
work out just as in GR to give
\beq H = \frac{\dot{a}}{a} \eeq
\subsection{Deceleration Parameter}
We can find an identity for the deceleration parameter explicitly. We have that
\beq trp \equiv \text{constant} \eeq
that is
\beq a^{3}\dot{a} \equiv \text{constant} \eeq
Differentiating across and rearranging slightly we find that
\beq \ddot{a} \equiv - \frac{3\dot{a}^{2}}{a} \eeq
Substituting this into the formula for the deceleration parameter $q$ gives
\beq
\begin{split} q = & - \frac{\ddot{a}}{a}\frac{1}{H^{2}} \\ = & - 
\frac{\ddot{a}a}{\dot{a}^{2}} \\ = & - \frac{-3\dot{a}^{2}a}{a\dot{a}^{2}} \\ =
 & \;\;3 \end{split} \eeq
\section{Problems of the Standard Cosmology}
There are a number of well known problems with the standard cosmology of GR. 
How does the new theory stand up to these? Let's consider them each in turn.
\subsection{The Cosmological Constant Problem}
This is probably the best known problem. In GR we have the following. There is 
a discrepancy of at least $120$ orders of magnitude between the possible value 
of $\Lambda$ today and what is expected at the Planck epoch taking the 
interpretation of $\Lambda$ as a vacuum energy. In GR the cosmological constant
 appears with the scalar curvature in the form $R + \Lambda$. However, in the 
new theory here it appears with a volume coefficient in the form $R + 
\frac{\Lambda}{V^{2/3}}$. Now, let us consider the change in volume since the 
Planck epoch. The radius of the universe was approximately $10^{-35}$cm at the 
Planck epoch. Today, the radius of the universe is about $10^{28}$cm. Thus the 
ratio of the volume at the Planck epoch to the volume today is about
\beq \frac{V_{0}}{V_{\text{Pl}}} = \biggl(\frac{10^{28}}{10^{-35}}\biggr)^{3} =
 10^{189} \eeq
Thus
\beq \frac{V^{2/3}_{0}}{V^{2/3}_{\text{Pl}}} = 10^{126} \eeq
If we were to consider the quantity $\frac{\Lambda}{V^{2/3}}$ as the 
cosmological ``constant'' we would have the problem encountered in GR. However,
the recognition of $\Lambda$ as the constant and recognising the presence of 
the volume term removes any problem whatsoever from this theory.
\\ \\
We should point out that this is fundamentally different from postulating a 
``time-varying cosmological constant''. The constant enters at the same level 
in his theory as in GR and it is the behaviour of the scalar curvature which 
changes things.
\subsection{The Flatness Problem}
The Flatness Problem In GR is entirely a product of the Hamiltonian constraint.
In GR the Hamiltonian constraint determines the dynamics and different energy 
values change the dynamics. However, in new CG, the Hamiltonian 
constraint is purely an identity and the flatness problem simply doesn't exist.
 The universe expands eternally according to the conformal constraint. This 
also has implications for the notion of dark matter. Although dark matter is 
sometimes employed to explain the otherwise strange behaviour of particular 
systems, one of the major reasons is to explain the apparent lack of matter 
needed to provide the observed flatness of the universe. The conformal theory 
needs no such strange explanations.
\subsection{The Horizon Problem}
To address this problem let's do a more in-depth analysis. To begin with let's 
return to GR and discuss the problem.
\subsubsection{Arc Parameter Time}
The ($4-$dimensional) FRW metric for the closed universe is
\beq ds^{2} = -dt^{2} + a(t)^{2}\biggl[d\chi^{2} + 
\text{sin}^{2}\chi(d\theta^{2} + \text{sin}^{2}\theta d\phi^{2})\biggr] \eeq
Of course, we can reparameterise this as
\beq ds^{2} = a(t)^{2}\biggl(d\eta^{2} + d\chi^{2} + 
\text{sin}^{2}\chi(d\theta^{2} + \text{sin}^{2}\theta d\phi^{2})\biggr) \eeq
The parameter $\eta$ is the \emph{arc parameter time}. It is a measure of the 
distance travelled by a photon along the surface of the three-sphere. In GR we 
can get values for two different times in particular. The Hubble time $H^{-1}$ 
and the proper time $t$ since the beginning of expansion. The Hubble time in 
terms of $\eta$ is
\beq H^{-1} = \frac{a(\eta)^{2}}{da/d\eta}^{2} \eeq
This is given by \cite{mtw}
\beq H^{-1} = \frac{a_{\text{max}}}{2}\frac{(1 - 
\text{cos}\eta)^{2}}{\text{sin}\eta} \eeq
where
\beq a_{\text{max}} = \frac{8\pi}{3}\rho_{m_{0}}a_{o}^{3} \eeq
The actual time since the beginning of expansion is given by \cite{mtw}
\beq t = \frac{a_{\text{max}}}{2}(1 - \text{cos}\eta) \eeq
If we approximate these numbers as
\beq H^{-1} = 20 \times 10^{9} \text{lyr} \eeq
and
\beq t = 10 \times 10^{9} \text{lyr} \eeq
then we can find the total distance travelled by a photon travelling on the 
three sphere since the very start of expansion
\beq
\begin{split} \frac{20 \times 10^{9} \text{lyr}}{10 \times 10^{9} \text{lyr}} =
 & \frac{H^{-1}}{t} \\ = & \frac{\frac{a_{\text{max}}}{2}\frac{(1 - 
\text{cos}\eta)^{2}}{\text{sin}\eta}}{\frac{a_{\text{max}}}{2}(1 - 
\text{cos}\eta)} \end{split} \eeq
From this we find that
\beq \eta = 1.975^{\text{o}} \eeq
Thus the horizon size at decoupling is only about $2^{\text{o}}$. However, 
there are about $10^{5}$ different regions of this size in the cosmic microwave
 background (CMB) sky. That is, about $10^{5}$ causally disconnected regions 
and yet we observe isotropy to about one part in $10^{5}$. This is the horizon 
problem in GR.
\\ \\
What can we discover about this from the conformal theory? We need to find 
$a(\eta)$ and $t(\eta)$. We found earlier the equation for $a(t)$ 
(\ref{at1}). This was 
\beq a(t)^{4} = Ct + a_{i}^{4} \eeq
where $C$ was a constant(spatially and temporally) and $a_{i}$ is the radius at
time $t = 0$. Let's take $a_{i}$ to be zero. Thus
\beq a(t)^{4} = Ct \eeq
now
\beq \eta = \int\frac{dt}{a(t)} \eeq
Solving this we get
\beq \eta = \frac{4}{3C^{1/4}}t^{3/4} \eeq
Inverting this to get $t(\eta)$ gives
\beq t = \biggl(\frac{4}{3}\biggr)^{4/3}C^{1/3}\eta^{4/3} \eeq
Then we find $a(\eta)$ easily
\beq a(\eta) = C^{1/4}\biggl(\frac{3}{4}\biggr)^{1/3}C^{1/12}\eta^{1/3} \eeq
Differentiating we get
\beq \frac{da}{d\eta} = 
\frac{1}{3}C^{1/3}\biggl(\frac{3}{4}\biggr)^{1/3}\eta^{-2/3} \eeq
Then
\beq H^{-1} = \frac{a^{2}}{da/d\eta} = 
3C^{1/3}\biggl(\frac{3}{4}\biggr)^{1/3}\eta^{4/3} \eeq
Finding the same ratio of $H^{-1}$ and $t$ as in GR we get
\beq \frac{H^{-1}}{t} = \frac{3C^{1/3}\biggl(3/4\biggr)^{1/3}
\eta^{4/3}}{\biggl(3/4\biggr)^{4/3}C^{1/3}\eta^{4/3}} \eeq
This simplifies very easily to give just
\beq \label{ugh} \frac{H^{-1}}{t} = 4 \eeq
There is no dependence on $\eta$ whatsoever!
\\ \\
Indeed, if we look at either of the expressions for $H^{-1}$ and $t$ in terms 
of $\eta$ and demand that the distance travelled by a photon since the 
beginning of expansion satisfy the seeming experimental value then what we get 
is a bound on the constant $C$ which in turn gives a bound on $trp$ which is 
already bounded by the values of $\Lambda$, $\rho_{d}$ and $\rho_{m}$. There is
 no horizon problem if the figures match up!! Of course, this removes the need 
to look for things like inflation which may in turn throw up a problem of its 
own in looking for large-scale structure formation. At first thought however, 
it does not seem like there will be a significant problem.
\\ \\
Another speculative idea which would require further work is one which could 
give rise to inflation. If at some stage in the past the conformal symmetry 
were to be broken we could envisage a situation where the then physical field 
$\psi$ might take on the role of the inflaton before decaying to the purely 
gauge field we treat in the theory. Again, I must stress that this is pure 
speculation rather than the ``solid'' prediction of a crank...
\section{Some Numbers}
We can actually make some concrete calculations very easily. The ratio of the 
cosmological constant at the Planck epoch to that of today is
\beq \frac{\Lambda_{\text{Pl}}}{\Lambda_{0}} = 10^{121} \eeq
at the very least. This can give us the ratio of the volume of the universe 
today to that of the Planck epoch
\beq \biggl(\frac{V_{0}}{V_{\text{Pl}}}\biggr)^{2/3} = 10^{121} \eeq
and from this the ratio of the radius today to the Planck length
\beq \frac{R_{0}}{R_{\text{Pl}}} = \sqrt{10^{121}} = 3.2 \times 10^{60} \eeq
where $R_{\text{Pl}}$ is the Planck length $1.7 \times 10^{-35}$m. Thus we get
\beq R_{0} = 5.4 \times 10^{25}\text{m} \eeq
for the minimum radius of the universe today.
\\ \\
The relationship found in (\ref{ugh})
\beq t = \frac{1}{4H} \eeq
can be used to get an estimate for the age of the universe. Taking the Hubble 
constant today to be
\beq H_{0} = 60 \text{km}/\text{s}/\text{Mpc} = 1.9 \times 10^{-18} 
\text{s}^{-1} \eeq
gives
\beq t_{0} = 1.32 \times 10^{17} \text{s} \eeq
A quick check of the value of $ct_{0}$ gives
\beq ct_{0} = 4 \times 10^{25} \text{m} \eeq
which is very close to our value of $R_{0} = 5.4 \times 10^{25}\text{m}$. The 
age of the universe is predicted from other means to be approximately $4.4 
\times 10^{17} \text{s}$. This is about $3$ times our value. All the same, we 
have found it from very elementary reasoning and to exactly the same order. One
 point here is that we found the deceleration parameter $q_{0}$ to be exactly 
$3$. This is higher than expected. However, $q_{0}$ is notoriously difficult to
 measure and perhaps in the light of the predictions here it should be 
re-examined. If indeed the commonly used value of $q_{0}$ (about $1.5$) were 
too low than the value of $H_{0}$ would be too high. Then our value of $t_{0}$ 
would go up by a factor similar to the correcting factor for $q_{0}$. Thus we 
see how a factor of $3$ might arise.
\\ \\
Suppose for a while that $H_{0}$ is indeed lower than accepted. Consider the 
following expression
\beq H_{0} = \frac{c}{d_{\text{L}}}f(q_{0},z) \eeq
where $f(q_{0},z)$ is a function of $q_{0}$ and $z$ (the redshift) only. From 
this we see that a large $H_{0}$ acts like a small $d_{\text{L}}$. Thus, if our
value of $H_{0}$ is too large we might interpret it as saying that 
$d_{\text{L}}$ is too small. A smaller than expected $d_{\text{L}}$ is exactly 
what is found in the recent supernovae experiments resulting in an apparent 
acceleration of the universe. The higher value of $q_{0}$ might possibly 
reconcile this with the theory here as the function $f$ behaves very roughly 
like $q_{0}^{-1}$.
\\ \\
Of course, a more obvious explanation may be that the simple FRW metric is just
not a perfect model for the universe. Perhaps we need a non-standard cosmology.
\section{A Non-Standard Cosmology: Anisotropy}
The standard FRW universes are all homogeneous and isotropic. What, if 
anything, does the new theory say about the subject of anisotropy. One obvious 
anisotropic model we can examine quite easily is the Kasner model. 
\subsection{The Kasner Universe}
The Kasner metric is
\beq d\sigma^{2} = t^{2p_{1}}dx^{2} + t^{2p_{2}}dy^{2} + t^{2p_{3}}dz^{2} \eeq
where 
\beq p_{1} + p_{2} + p_{3} = (p_{1})^{2} + (p_{2})^{2} + (p_{3})^{2} = 1 \eeq
Each $t = $ constant hypersurface of this cosmological model is a flat 
three-dimensional space. It represents an expanding universe since
\beq \sqrt{g} = t \eeq
is constantly increasing. However, its expansion is anisotropic. Consider two 
standard observers. If only their $x-$coordinates differ than their separation 
is given by $t^{2p_{1}}\Delta x$. Thus distances parallel to the $x-$axis 
expand at one rate $l_{1} \varpropto t^{p_{1}}$ while those along the $y-$axis 
expand at a different rate $l_{2} \varpropto t^{p_{2}}$. A truly remarkable 
feature is that along one of the axes distances contract rather than expand. 
This is because one of the $p_{i}$ must be non-positive. Let's calculate the 
extrinsic curvature of this model. This is very straightforward. We find that
\beq K_{ii} = - 2p_{i}t^{2p_{i}-1} \eeq
Thus we get
\beq
\begin{split} trK = & \Sigma _{i}t^{-2p_{i}}(-2p_{i}t^{2p_{i}-1}) \\ = & - 
2t^{-1}\Sigma p_{i} \\ = & - 2t^{-1} \end{split} \eeq
If the universe we are considering is not closed then we know that
\beq trp = 2trK \eeq
and here that means that $trp$ is time-dependent which of course is not allowed
 by the theory. Thus we \emph{cannot} have a non-closed Kasner model. Let's try
now the same metric but where the $x$, $y$ and $z$ coordinates are interpreted 
as angles with period $4\pi$. This model \emph{is} closed. The volume of a 
hypersurface is given by
\beq
\begin{split} V = & \int\sqrt{g}\;\;d^{3}x \\ = & \int t\;\;d^{3}x \\ = & 
t(xyz)|^{4\pi}_{0} \\ = & 64t\pi^{3} \end{split} \eeq
The form of $trK$ is unchanged but now
\beq trp = 2trKV^{4/3} \eeq
and so we have
\beq trp = 1024\pi^{4}t^{1/3} \eeq
However, this is still a time-varying quantity! Thus the Kasner model also 
seems a lost cause.
\subsection{Effective Anisotropic Energy Density}
The physics of the anisotropic scenario can be discussed in other language. In 
GR the idea is to write the Hamiltonian constraint as
\beq \pi^{ab}\pi_{ab} - \frac{1}{2}(tr\pi)^{2} -g(R - \rho_{\text{m}} - 
\rho_{\text{an}}) = 0 \eeq
where $\rho_{\text{an}}$ is the anisotropy energy density and $\rho_{{m}}$ is 
the energy density of matter (what ever that matter happens to be). The 
anisotropy energy density is found to have an equation of state
\beq \rho_{\text{an}} \varpropto g^{-1} \eeq
That is
\beq \rho_{\text{an}} \varpropto V^{-2} \eeq
Thus we can write
\beq \rho_{\text{an}} = \rho_{\text{an}_{0}}\frac{V_{0}^{2}}{V^{2}} \eeq
where the subscript $0$ means the value today. In the conformal Hamiltonian 
constraint this appears with $R$ as
\beq \frac{R}{V^{4/3}} - \rho_{\text{an}_{0}}\frac{V_{0}^{2}}{V^{2}} = 
\frac{1}{V^{4/3}}\biggl(R - 
\rho_{\text{an}_{0}}\frac{V_{0}^{2}}{V^{2/3}}\biggr) \eeq
Thus we get the characteristic $V^{-2/3}$ factor even for the anisotropy which 
is encouraging.
\section{Discussion}
The cosmological scenario presented here is very interesting. Not alone have we
 managed to recover expansion (our initial inspiration for the new theory) but 
we have found many other desirable and exciting features. Among these are the 
following: the theory seems incredibly restrictive: we are forbidden to have an
 open or flat universe (at least of the FRW type); we have specific bounds on 
the various matter sources in terms of each other; some of the major problems 
of the standard GR cosmology have been resolved (and some that haven't may yet 
succumb to this theory); we have a definite prediction for the deceleration 
parameter; we are restricted in the types of anisotropy we may have. Such 
successes at such an early stage are promising and further work must surely be
 warranted.
\chapter{Discussion}
The path of general relativity has led from Einstein's spacetime to the 
$($space $+$ time$)$ of Dirac and ADM, to Wheeler's superspace and to finally 
to York's conformal superspace. Rather than taking this route we place 
conformal superspace in the central position to begin with and find a theory 
which gives a lot while taking very little.
\\ \\
Of course, one might argue that just because this theory is self consistent is 
not reason enough to demand further attention. However, a recent result of 
\'{O} Murchadha \cite{nom} shows that if we demand a constrained Hamiltonian 
with a closed constraint algebra then we are severely limited in our options. 
In fact, there are essentially only $4$ options. These are
\\ \\
(i) Regular GR \\
(ii) A maximally sliced theory \\
(iii) A constant mean curvature sliced theory \\
(iv) Strong gravity
\\ \\
The first and fourth are known and exist in their own right. However, the 
second and third are seen to arise naturally in the conformal approach adopted 
here. (In fact, we could consider option (ii) to be a special case of option 
(iii) and see the options decrease even further.) We have found a consistent 
theory and since these are so rare, it is surely a worthy result in its own 
right.
\\ \\
Of course, experiment will always have the last word and any theory is only as 
good as its predictions. How does our theory stand up to experiment? The solar 
system results seem to hold and this is also expected to be true for the binary
 pulsar. However, it is in cosmology that the beauty of this theory is most 
apparent.
\\ \\
The theory seems to accept very few cosmological solutions. It is incredibly 
restrictive. We are denied the FRW open and flat universes. We are also denied 
the Kasner universe. The fact that $trK$ is (spatially) constant places a very 
severe restriction on what anisotropy (if any) is possible. The constant $trK$ 
is a property of the theory itself and not simply a property of a particular 
cosmological solution. The Hamiltonian constraint is elevated to an identity 
and the dynamics are found in the conformal constraint $trp = \text{constant}$.
\\ \\
Most of the problems of the standard cosmology of GR do not occur and in fact 
there is  a possibility that \emph{all} of the problems may be removed. Of 
course, there is also the chance that the predictions of the deceleration 
parameter and the age of the universe may prove to be the downfall of the 
theory. The fact that these predictions are so easy to find however is a 
positive thing and as they say, ``hope springs eternal.''
\\ \\
From a quantisation view point the theory has several attractive features. 
Firstly, with regard to the static theory, the absence of $tr\pi$ removes 
various problems since the quantity $\pi^{ab}\pi_{ab}$ is positive definite 
unlike the quantity $\pi^{ab}\pi_{ab} - \frac{1}{2}(tr\pi)^{2}$ in GR. The fact
 that the configuration space is smaller is also attractive. While this theory 
seems unlikely to be a good model of reality it may nonetheless teach some 
valuable lessons with regard to the quantisation of gravity.
\\ \\
The non-static theory is also attractive from a quantisation point of view. We 
have a similar advantage with the reduced configuration space. The most attractive feature may be the emergence of a physically preferred slicing. The problem
 of time is the major stumbling block in the path to a quantisation of gravity 
and this preferred slicing may prove to be invaluable. We also have an added 
bonus in that the volume of the universe is monotonically increasing from $0$ 
to $\infty$ and is trivially constant on any hypersurface. Thus the volume may 
be of use as a notion of time in the theory.
\\ \\
As regards cosmology again, the elevation to identity of the Hamiltonian 
constraint may be crucial. The constant $trp$ constraint is a \emph{far} more 
elementary quantity and this may be of no-small help with regard to quantising 
the cosmological solution.
\\ \\
All this indicates that a quantisation program for the theory would be 
beneficial regardless of the eventual fate of the theory as a classical 
competitor to GR. Of course, if that fate were to be a positive one then I for
 one won't be complaining...

\end{document}